\documentclass[prd,preprint,superscriptaddress,preprintnumbers,eqsecnum,showpacs,nofootinbib,nobibnotes]{revtex4}
\usepackage{amsfonts,amsmath,amssymb,bm,natbib}
\usepackage{graphicx} 
\usepackage{color}

\newcommand{\be}{\begin{equation}}
\newcommand{\bea}{\begin{eqnarray}}
\newcommand{\ee}{\end{equation}}
\newcommand{\eea}{\end{eqnarray}}

\def\G1{\widetilde{\Gamma_{1}}}

\def\s#1{{\scriptscriptstyle #1}}

\def\noeq#1{(\ref{#1})}
\def\1eq#1{Eq.~(\ref{#1})}

\def\2eqs#1#2{Eqs.~(\ref{#1}) and~(\ref{#2})}
\def\3eqs#1#2#3{Eqs.~(\ref{#1}),~(\ref{#2}) and~(\ref{#3})}

\def\fig#1{Fig.~\ref{#1}}

\def\ie{{\it i.e.}, }
\def\eg{{\it e.g.}, }

\def\n#1{({\it #1}\,)}

\def\gnp{\Gamma^{ \bf np}}

\def\gp{\Gamma^{ \bf p}}

\def\d{(q\spr t)r}

\def\d{{\mathrm{d}}}

\begin{document}

\title{Schwinger mechanism in linear covariant gauges}


\author{A.~C. Aguilar}
\affiliation{University of Campinas - UNICAMP, 
Institute of Physics ``Gleb Wataghin'',
13083-859 Campinas, SP, Brazil}

\author{D. Binosi}
\affiliation{European Centre for Theoretical Studies in Nuclear
Physics and Related Areas (ECT*) and Fondazione Bruno Kessler, \\Villa Tambosi, Strada delle
Tabarelle 286, 
I-38123 Villazzano (TN)  Italy}

\author{J. Papavassiliou}
\affiliation{\mbox{Department of Theoretical Physics and IFIC, 
University of Valencia and CSIC},
E-46100, Valencia, Spain}

\begin{abstract}

In this work we explore the applicability of a special gluon mass generating mechanism in the context of the linear covariant gauges.
In particular, the implementation of the 
Schwinger mechanism in  pure Yang-Mills theories hinges crucially on the inclusion  
of massless bound-state excitations in the fundamental nonperturbative vertices 
of the theory. The dynamical formation of such excitations 
is controlled by a homogeneous linear Bethe-Salpeter equation, 
whose nontrivial solutions have been studied only in the Landau gauge.
Here, the form of this integral equation is derived for general values of the gauge-fixing parameter, under 
a number of simplifying assumptions that reduce the degree of technical complexity. 
The kernel of this equation consists of fully-dressed gluon propagators, 
for which recent lattice data are used as input,  and of three-gluon vertices dressed by a single form factor,
which is modelled by means of certain physically motivated Ans\"atze. 
The gauge-dependent terms contributing to this kernel 
impose considerable restrictions on the infrared behavior of the vertex form factor; specifically,    
only infrared finite Ans\"atze are 
compatible with the existence of nontrivial solutions. When such Ans\"atze are employed, 
the numerical study of the integral equation  reveals a continuity in the type of solutions 
as one varies the gauge-fixing parameter, 
indicating a smooth departure from the Landau gauge. Instead, the logarithmically divergent form factor 
displaying the characteristic ``zero crossing'', while perfectly consistent in the Landau gauge,  
has to undergo a dramatic qualitative transformation away from it, in order to yield acceptable 
solutions. The possible implications of these results are 
briefly discussed.

\end{abstract}


\maketitle

\section{\label{int}Introduction}                                                      

In recent years, numerous large-volume latice simulations in the Landau gauge, both for SU(2)~\cite{Cucchieri:2007md,Cucchieri:2007rg,Cucchieri:2009zt,Cucchieri:2010xr} and SU(3)~\cite{Bogolubsky:2009dc,Bogolubsky:2007ud,Bowman:2007du,Oliveira:2009eh}, have firmly established that the scalar form factor of the gluon propagator, to be denoted by $\Delta(q^2)$, saturates at a finite (nonvanishing) value in the deep infrared (IR), \ie  $\Delta(0) = c >0$. Even though this particular theoretical possibility had been anticipated in a variety of works spanning several decades~\cite{Smit:1974je,Cornwall:1979hz,Cornwall:1981zr,Bernard:1981pg,Bernard:1982my,Donoghue:1983fy,Lavelle:1991ve,Halzen:1992vd,Leinweber:1998uu,Philipsen:2001ip,Szczepaniak:2001rg,Aguilar:2001zy,Aguilar:2002tc,Maris:2003vk,Szczepaniak:2003ve,Li:2004te,Aguilar:2004sw}, the paradigm-shifting nature of this result sparked an intense activity among QCD practitioners, and several distinct mechanisms have been put forth in order to explain it, and explore its relation with other fundamental phenomena, such as confinement, chiral symmetry breaking, and hadron formation~\cite{Aguilar:2006gr,Kondo:2006ih,Braun:2007bx,Epple:2007ut,Binosi:2007pi,Binosi:2008qk,Aguilar:2008xm,Boucaud:2008ky,Dudal:2008sp,Fischer:2008uz,Binosi:2009qm,Szczepaniak:2010fe,Watson:2010cn,RodriguezQuintero:2010wy,Campagnari:2010wc,Tissier:2010ts,Aguilar:2010cn,Pennington:2011xs,Watson:2011kv,Kondo:2011ab,Qin:2011dd,Fister:2013bh,Cloet:2013jya,
	Binosi:2014aea,Roberts:2015dea,Aguilar:2016vin}.
	
Of course, as is well-known, Green's functions depend in general on the gauge-fixing scheme used for the quantization of the theory, and the gauge-fixing parameter chosen within a given scheme. In one of the most commonly used gauge-fixing procedures, the linear covariant (or  $R_\xi$) gauges~\cite{Fujikawa:1972fe}, the corresponding term that must be added to the standard Yang-Mills Lagrangian is given by $\frac{1}{2\xi} (\partial^\mu A^a_\mu)^2$, where $\xi$ represents the gauge-fixing parameter; some characteristic values include the aforementioned Landau  gauge ($\xi=0$) and the Feynman gauge ($\xi=1$). Therefore, one important question that arises naturally in this context is whether the observed IR finiteness of the gluon propagator is particular to the Landau gauge, or whether it persists away from it. 

It turns out that recent studies in this direction indicate that this particular property does in fact survive even if  $\xi\neq 0$. At the level of lattice simulations, the implementation of a novel algorithm~\cite{Cucchieri:2009kk} revealed the same feature in gluon propagators with minute positive values of $\xi$~\cite{Cucchieri:2011pp}. A stronger indication was found in the simulations of~\cite{Bicudo:2015rma}, where the IR finiteness of the gluon propagators was confirmed\footnote{To be sure, the saturation point itself is not fixed, but varies as a function of $\xi$, decreasing as  $\xi$ increases. What seems to be $\xi$-independent, however, is the qualitative property of IR saturation at a nonvanishing value.} for  values of the $\xi$ ranging within the interval~[0,0.5]. In addition, in~\cite{Aguilar:2015nqa} was argued that the Nielsen identities~\cite{Nielsen:1975fs} support this general picture, but no particular statement was made regarding the explicit influence of the gauge-fixing parameter on the underlying dynamics. Finally, massive propagators for general $\xi$ have been derived in~\cite{Capri:2015ixa,Capri:2015nzw}, within the refined  Gribov-Zwanziger framework~\cite{Dudal:2008sp}. 

In a series of articles~\cite{Aguilar:2011xe,Ibanez:2012zk,Aguilar:2016vin}, a particular framework for the study of the IR finiteness of the gluon propagator has been elaborated, which constitutes a particular realization of the Schwinger mechanism in a non-Abelian context ~\cite{Schwinger:1962tn,Schwinger:1962tp,Jackiw:1973tr,Jackiw:1973ha,Smit:1974je,Eichten:1974et,Poggio:1974qs}. A central ingredient of this approach is the dynamical formation of longitudinally coupled massless bound-state excitations,  whose main effect is to introduce poles in the nonperturbative vertices of the theory. The inclusion of these poles enables the evasion of a powerful {\it $\xi$-independent} cancellation operating at the level of the Schwinger-Dyson equation (SDE) 
for $\Delta(q^2)$, which would have otherwise forced  the {\it  exact} vanishing of  $\Delta^{-1}(0)$~\cite{Aguilar:2016vin}. Thus, in the context of this particular mechanism, the gluon mass generation, \ie the property $\Delta^{-1}(0) =m^2$, is intimately connected with the ability of the theory to create such massless poles dynamically. Their actual formation is governed by a homogeneous Bethe-Salpeter equation (BSE), whose solutions determine the so-called ``bound-state wave function'': if the BSE admits nontrivial solutions, the mass generation mechanism is activated; in the contrary case of obtaining only an identically vanishing solution, the formation of such bound-state excitations is not dynamically realized, and no gluon mass can be possibly generated. 

This  particular  BSE  has   been  solved  under  certain  simplifying
approximations  only in  the  Landau gauge,  where  the generation  of
nontrivial          solutions         has          indeed         been
established~\cite{Aguilar:2011xe,Ibanez:2012zk}; however, no analogous
study has  ever been carried out  for $\xi\neq 0$. The  purpose of the
present work is  to provide the first preliminary  exploration of this
important theoretical  issue. In particular,  in what follows  we will
derive  the  BSE  for  general values  of  $\xi$ in the context of a pure Yang-Mills theory (no quarks),  and  
then analyze under what conditions we may 
obtain from it nontrivial  solutions, at least 
for the  values of~$\xi$  within the interval \mbox{[0, 0.5.]}, used in
the lattice simulations of~\cite{Bicudo:2015rma}.
 
From the purely  conceptual point of view, the steps  that lead to the
BSE in  question, as well  as the connection  of its solutions  to the
value of $\Delta^{-1}(0)$,  do not depend on the  particular choice of
the gauge-fixing  parameter; what changes  with respect to  the Landau
gauge is rather the explicit form  of the kernel appearing in the BSE.
This kernel is in general composed of full gluon propagators [see~\1eq{QQprop}] and 
three-gluon vertices; the latter are assumed to be proportional to their 
tree-level tensors, and are multiplied by a global form factor, to be denoted by $f$, 
which ``dresses'' them with nonperturbative effects [see~\1eq{vertf}]. 
It turns out that away from the  Landau gauge a proliferation of terms
is produced,  proportional to various  powers of $\xi$,  stemming from
the nontransverse  parts of the  gluon propagators entering into the kernel. 
The effect of these new terms is not 
only quantitative, in the sense of giving rise to more complicated algebraic expressions, 
but also qualitative, given that one of them turns out to be more divergent in the IR  
than all the others. This fact may be understood 
by noting that, whereas the Landau gauge parts contribute to the kernel  
an effectively massive propagator (\ie the $\Delta(k^2)$, {\it assuming that the mass is indeed generated}), 
the corresponding $\xi$-dependent parts 
furnish a massless one (\ie $1/k^2$, {\it even if the mass is generated}). As we will see,  
the accumulation of such ``massless propagators'' is more acute in the term with the maximum power of $\xi$, namely $\xi^3$, 
which becomes potentially ``unstable''. 
Specifically, when the  $f$ employed is finite in the IR, this particular term 
does not affect qualitatively the situation; however, when $f$ diverges logarithmically (as indicated by a variety of recent studies~\cite{Aguilar:2013vaa,Pelaez:2013cpa,Blum:2014gna,Eichmann:2014xya,Cyrol:2016tym,Athenodorou:2016oyh,Duarte:2016ieu}), the form of the $\xi^3$ term ``destabilizes'' the solution of the BSE.

Given the above observations, we have 
evaluated the BSE kernel using as input the $\Delta(q^2,\xi)$
from the lattice~\cite{Bicudo:2015rma}, but have employed qualitatively 
distinct Ans\"atze for $f$. In particular, we used \n{i} two functional forms for $f$, 
sharing the common characteristic of being finite at the origin, and 
\n{ii} an Ansatz for $f$ that reverses sign in the IR (``zero-crossing'') and diverges logarithmically 
as it approaches the origin.

The main conclusions that can be  drawn from the analysis of these two cases are completely different. 

In case \n{i},  the changes induced to the BSE by 
the fact that the $\Delta(q^2,\xi)$ become gradually suppressed as $\xi$ increases~\cite{Bicudo:2015rma}, 
 may be ``reabsorbed'' into mild modifications in the form of $f$, such that   
eventually a solution for each  value of $\xi$ within the interval [0, 0.5.] may be found. 
In  that sense, the departure from the Landau gauge  is rather smooth  and completely  stable,
and the term proportional to $\xi^3$ remains under control. 

The situation in case \n{ii} is far more intriguing. Plainly, the IR divergence of $f$, 
combined with the destabilizing tendency of the $\xi^3$ term, prevents the 
BSE from having a nontrivial solution {\it away} from the Landau gauge. 
To overcome this difficulty, one has to postulate that 
the logarithmic divergence of $f$ is very particular to the Landau gauge, 
and, as one departs from it, $f$ reaches negative but {\it finite} values at the origin.
After imposing this special assumption on $f$, one finds again nontrivial solutions 
for the BSE, which vary mildly as functions of $\xi$, and are very similar to that
obtained in the  Landau gauge (with the divergent $f$).

The  article  is organized  as  follows.  In Sect.~\ref{sec:genfr}  we
present  a brief  introduction  to the  main  concepts underlying  the
present work, with particular emphasis on the notions related with the
bound-state  poles. Then,  in Sect.~\ref{sec:masseq0},  we derive  the
general  expression  of  the transition  amplitude,  an  indispensable
ingredient  of  the  present  approach,   for  general  value  of  the
gauge-fixing parameter~$\xi$. In  Sect.~\ref{sec:bse} we undertake the
rather laborious task of deriving the BSE that controls the generation
of  the massless  bound-state excitations,  using certain  simplifying
assumptions regarding  the structure  of its  kernel. Even  though the
calculation is carried  out for a general~$\xi$, a  particular type of
contributions is eventually neglected, in order to reduce the level of
technical complexity.  Next, in  Sect.~\ref{sec:numan} we carry  out a
detailed  numerical  analysis  of  the BSE  derived  in  the  previous
section,  using  a  variety  of Ans\"atze  for the vertex form  factors
appearing in its  kernel, and focusing on the conditions  that must be
satisfied  in  order  to  obtain  nontrivial  solutions. Then, 
in  Sect.~\ref{sec:glueball} we discuss the similarities and differences
between the BSEs studied here and those associated with the formation of glueballs. Finally,  in
Sect.~\ref{sec:c} we present our conclusions.

\section{\label{sec:genfr}General framework}

In this section we introduce the necessary notation and conventions, and review the main features of the general theoretical framework that will be employed throughout this work.

The gluon propagator in the $R_\xi$ gauges is given by (we suppress the color factor $\delta^{ab}$) 
\begin{align} 
\Delta_{\mu\nu}(q) = -i\left[\Delta(q^2)P_{\mu\nu}(q) + \xi\frac{q_\mu q_\nu}{q^4}\right]; \qquad P_{\mu\nu}(q) = g_{\mu\nu}-\frac{q_\mu q_\nu}{q^2},
\label{QQprop}
\end{align}
with inverse
\begin{align}
	\Delta_{\nu\rho}^{-1}(q) &= i\left[\Delta^{-1}(q^2)P_{\nu\rho}(q) + \xi^{-1}q_\nu q_\rho\right].
\label{invQQprop}
\end{align} 
The function $\Delta(q^2)$ is related to the gluon self-energy $\Pi_{\mu\nu}(q)=P_{\mu\nu}(q)\Pi(q^2)$  through $\Delta^{-1}({q^2})=q^2+i\Pi(q^2)$. Note that $\Pi(q^2)$, and therefore also $\Delta(q^2)$, depend explicitly on the value of $\xi$, \ie $\Delta(q^2) =  \Delta(q^2,\xi)$; in what follows this dependence will be displayed only when it is deemed to be conceptually advantageous.

\begin{figure}[!t]
\includegraphics[scale=0.625]{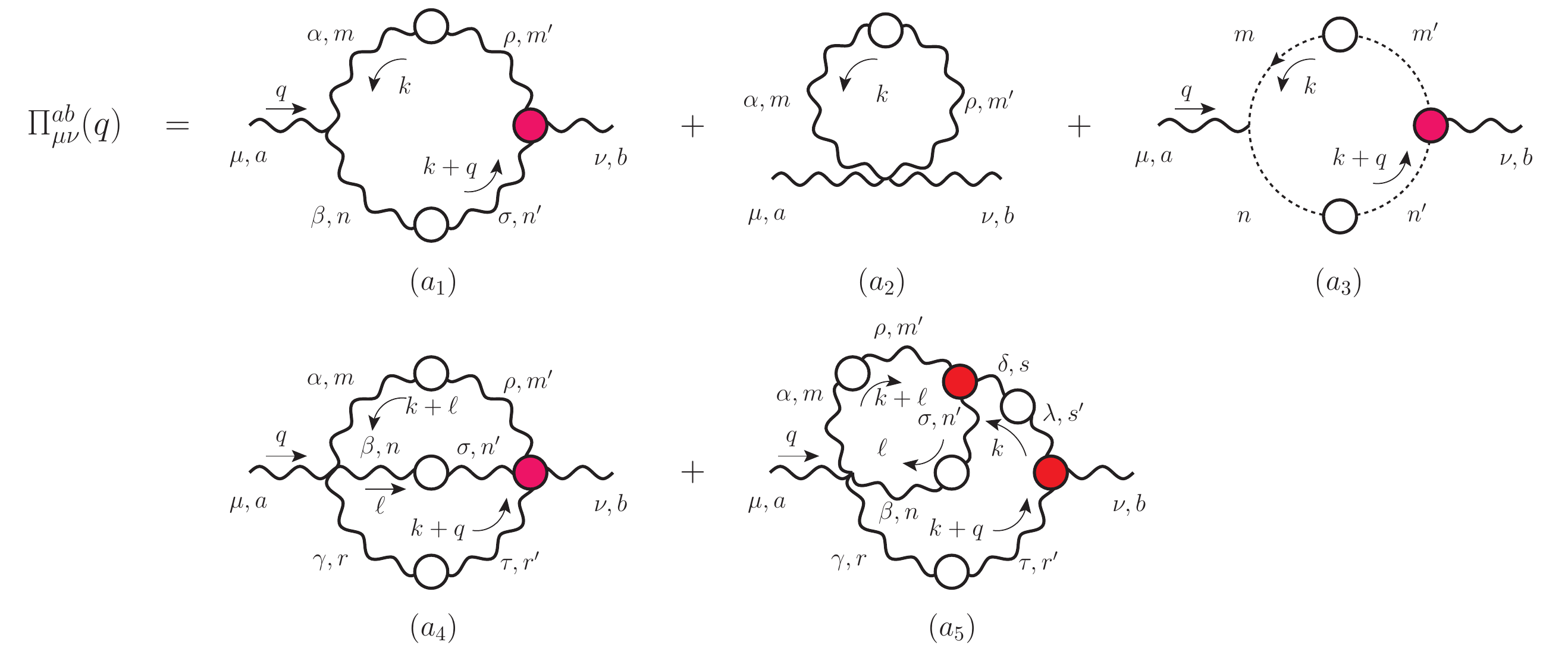}
\caption{\label{QQ-SDE}The standard SDE for the gluon self-energy in the
absence of quarks (pure Yang-Mills theory). White (colored) circles denote fully-dressed propagators (vertices).}
\end{figure}

Let us next consider the conventional SDE of the gluon propagator, valid for any value of the gauge-fixing parameter, namely  
\be
\Delta^{-1}(q^2) P_{\mu\nu}(q) = q^2P_{\mu\nu}(q)+ i\sum_{i=1}^5 (a_i)_{\mu\nu},
\label{QQSDE}
\ee
where the diagrams $(a_i)$ are shown in~\fig{QQ-SDE}.

In what follows we concentrate on the type of solutions of this SDE that display the characteristic feature of IR saturation, which may be interpreted as the result of the nonperturbative generation of an effective gluon mass. As has been shown in a series of previous works~\cite{Smit:1974je,Cornwall:1981zr,Aguilar:2008xm,Aguilar:2016vin}, the existence of such solutions requires the inclusion of terms proportional to $1/q^2$ in the fully dressed vertices appearing in the diagrams of Fig.~\ref{QQ-SDE}. Even though in principle the three main vertices entering into the gluon SDE may contain such massless poles, in what follows we will restrict our attention to the case of the three-gluon vertex, ${\Gamma}_{\mu\alpha\beta}(q,r,p)$, with $q+r+p=0$. 

In particular, following~\cite{Aguilar:2016vin}, we will separate the vertex into two distinct parts, namely 
\be
{\Gamma}_{\mu\alpha\beta}(q,r,p) = \gnp_{\mu\alpha\beta}(q,r,p) + \gp_{\mu\alpha\beta}(q,r,p),
\label{npandp}
\ee 
where the superscripts `${\bf np}$' and `${\bf p}$' stand for `no-pole' parts and `pole' parts, respectively. The part $\gnp_{\mu\alpha\beta}$ may be expanded in the ``naive'' basis used in~\cite{Aguilar:2016vin}, or in the well-known ``longitudinal'' and ``transverse'' basis of Ball and Chiu~\cite{Ball:1980ax}. The important point is that, since the form factors multiplying the 14 possible tensors of either basis do not contain poles of the type  $1/q^2$, $\gnp_{\mu\alpha\beta}$ is ``inert'' as far as the mass generation is concerned. In fact, under exactly analogous assumptions for the form factors of the ghost-gluon and four-gluon vertices entering into the SDE of~\fig{QQ-SDE}, certain crucial identities are triggered, which enforce the exact relation\footnote{In~\cite{Aguilar:2016vin} the demonstration of this relation was carried out within the background-field method, where 
the right external leg entering into all fully-dressed vertices of \fig{QQ-SDE} is a background gluon. 
The advantage of the background field framework in this context is 
that the realization of the so-called ``seagull cancellations'' becomes 
far more transparent. Given that in the present work we are not interested 
in this particular aspect of the problem, and that the background and 
conventional Green's functions  are related 
by exact symmetry identities~\cite{Binosi:2013cea}, we have opted 
for the standard formulation of  Yang-Mills theory.}~$\Delta^{-1}(0) =0$.
 
\begin{figure}[!t]
\includegraphics[scale=0.625]{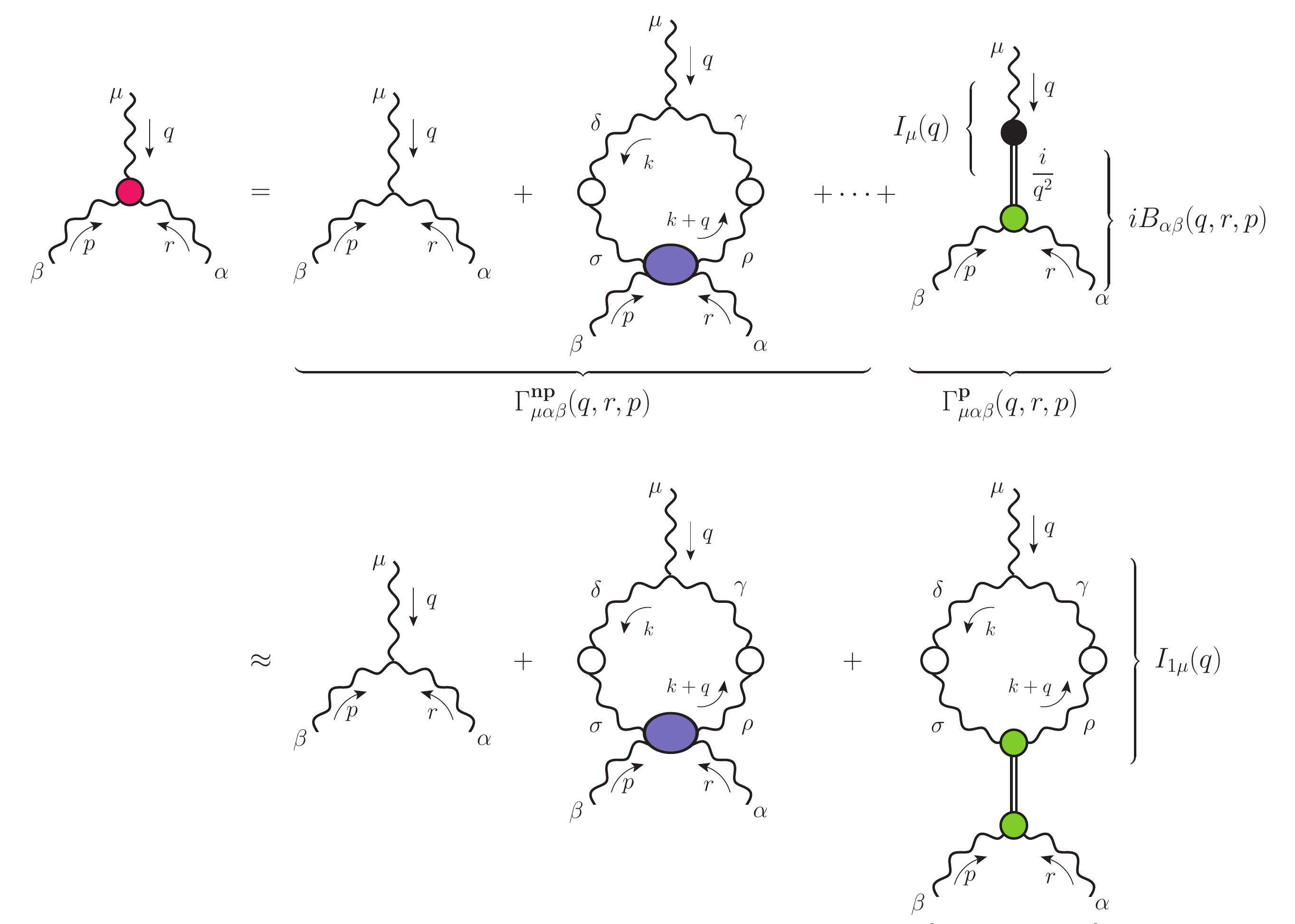}
\caption{\label{Gammaprime}The decomposition of the three-gluon vertex $\Gamma$ into its regular and pole part (first line), and the approximate version of this equation used in the paper (second line).}
\end{figure}
 
Turning to the component $\gp_{\mu\alpha\beta}$, since the origin of the poles contained in it is attributed to the formation of massless excitations, it is natural to employ the language of bound states in order to describe its structure. Specifically, $\gp_{\mu\alpha\beta}$ is composed of three fundamental ingredients, shown diagrammatically in~\fig{Gammaprime}: \n{i} The universal nonperturbative {\it transition amplitude}, to be denoted by ${I}_\mu(q)$,   which connects, in all possible ways, a single gluon to the massless excitation; \n{ii} the scalar massless excitation, whose propagator furnishes the pole ${i}/{q^2}$, and~\n{iii} the ``bound-state wave function'' (or ``proper vertex function''~\cite{Jackiw:1973ha}), to be denoted by $B_{\alpha\beta}$, which connects the massless excitation to the two gluons (carrying momentum $r$ and $p$).  Thus one has (suppressing the $f^{abc}$ on both sides)
\be
\gp_{\mu\alpha\beta}(q,r,p) = {I}_\mu(q) \times\frac{i}{q^2}\times i B_{\alpha\beta}(q,r,p).
\label{UIB}
\ee
Clearly, due to Lorentz invariance,
\be
{I}_\mu(q) = q_\mu {I}(q^2),
\label{Quantumtransition}
\ee
and $\gp_{\mu\alpha\beta}(q,r,p)$ is {\it longitudinally coupled} in the $q$-channel, as required~\cite{Jackiw:1973tr,Jackiw:1973ha,Smit:1974je,Eichten:1974et,Poggio:1974qs}.

The tensorial decomposition of the vertex $B_{\alpha\beta}(q,r,p)$ is given by  
\be
B_{\alpha\beta}(q,r,p) = B_1g_{\alpha\beta}+B_2 p_{\alpha}p_{\beta}+B_3r_{\alpha}r_{\beta}+B_4 r_{\alpha}p_{\beta} +B_5 p_{\alpha} r_{\beta},
\label{tensB}
\ee
where the form factors $B_i = B_i(q,r,p)$ are constrained by the Bose symmetry of $B_{\alpha\beta}$ with respect to the exchange of the two gluon fields. In particular, given that the color factor $f^{abc}$ has been factored out, Bose symmetry requires that, under the simultaneous exchange $\alpha \leftrightarrow \beta$ and $r \leftrightarrow p$, we must have
\be
B_{\alpha\beta}(q,r,p) = - B_{\beta\alpha}(q,p,r),
\label{BoseB}
\ee
which implies that 
\be
B_{\alpha\beta}(0,-p,p) = 0.
\label{B0}
\ee
At the level of the individual form factors, \1eq{BoseB} imposes the constraints
\begin{align}
B_i(q,p,r) &= - B_i(q,r,p) ; \qquad i=1,4,5 
\nonumber\\
B_2(q,p,r) &= - B_3(q,r,p),
\label{Bosecon}	
\end{align}
which imply that, at $q=0$, 
\begin{align}
B_i(0,-p,p) &= 0; \qquad i=1,4,5 
\nonumber\\
B_2(0,-p,p) &= -B_3(0,-p,p).
\label{Bose0}
\end{align}

Note that, at the formal level, all the aforementioned structural characteristics, known from the various analysis specialized in the Landau gauge~\cite{Aguilar:2011xe,Ibanez:2012zk}, are straightforwardly generalized to an arbitrary value of the gauge fixing parameter, and the nontrivial dependence on $\xi$ is completely implicit. However, as we will see in what follows, the various sources of $\xi$-dependence become evident as soon as the explicit evaluation of any of the above quantities is undertaken. 

It turns out that the saturation point of the gluon propagator, $\Delta(0)$, and the value  of the form factor of the transition amplitude at the origin, $I(0)$ are related by the simple formula
\begin{equation}
	\Delta^{-1}(0) = g^2 I^2(0), 
\label{mI}
\end{equation}
where $g$ is the gauge coupling.  This compact equation dates back to the pioneering work of~\cite{Eichten:1974et}, and its detailed derivation in the specialized context of gluon mass generation has been presented in~\cite{Ibanez:2012zk}.

In the following analysis we will simplify the situation by retaining only the contribution to ${I}_\mu(q)$ originating from the graph shown in the second line of~\fig{Gammaprime}, to be denoted by ${I}_{1\mu}(q)$, and its scalar form factor $I_{1}(q^2)$. Then, the {\it exact} relation given in \1eq{mI} is replaced by the {\it approximate} formula 
\begin{equation}
	\Delta^{-1}(0) \approx g^2 I_{1}^2(0), 
\label{mIapp}
\end{equation}
which will be employed in the present work.

\section{\label{sec:masseq0} The transition amplitude for general $\xi$}

In this section we consider the transition amplitude $I_{1\mu}(q)$, shown in~\fig{Gammaprime}, and derive a formula that expresses $I_{1}(0)$ in terms of the form factors $B_i$, for a general value of the gauge fixing parameter  $\xi$.

As already mentioned, Lorentz invariance implies that $I_{1\mu}(q)=q_\mu I_1(q)$, so that the scalar part of this amplitude is readily found to be 
\be
	I_{1}(q^2) = \frac{i}{2}C_A \frac{q^{\mu}}{q^2}\int_k\!\Gamma^{(0)}_{\mu\gamma\delta}(q,k,-q-k)\Delta^{\gamma\rho}(k)\Delta^{\delta\sigma}(q+k) {B}_{\sigma\rho}(q,-k-q,k),
\label{I1}
\ee
where we have introduced the dimensional regularization integral measure 
\be
\int_{k}\equiv\frac{\mu^{\epsilon}}{(2\pi)^{d}}\!\int\!\mathrm{d}^d k,
\label{dqd}
\ee
with $\mu$ the 't Hooft mass, and  $d=4-\epsilon$ the space-time dimension.

In order to obtain $I_{1}(0)$, one must carry out a Taylor expansion of the integrand around $q=0$, as the presence of the $q^2$ pole on the right-hand side (r.h.s.) of \1eq{I1} prevents its direct evaluation. In particular, making use of \1eq{B0},  
it is easy to establish that
\be 
	I_{1}(q^2) = \frac{i}{2}C_A  \frac{q^{\mu} q^\lambda}{q^2}\int_k\Gamma^{(0)}_{\mu\gamma\delta}(0,k,-k)\Delta^{\gamma\rho}(k)\Delta^{\delta\sigma}(k) \bigg\lbrace \frac{\partial {B}_{\sigma\rho}}{\partial q^\lambda}\bigg\rbrace_{\!\!q=0} + {\cal O}(q^2).
\label{a1C}
\ee
The integral appearing on the r.h.s of \1eq{a1C} has two free Lorentz indices, $\mu$  and $\lambda$, 
and no momentum scale; therefore, it can only be proportional to $g_{\mu\lambda}$, {\it i.e.}, 
\be 
	I_{1}(0)  = \frac{i}{2d}C_A  \int_k \Gamma^{(0)}_{\mu\gamma\delta}(0,k,-k)\Delta^{\gamma\rho}(k)\Delta^{\delta\sigma}(k) \bigg\lbrace \frac{\partial {B}_{\sigma\rho}}{\partial q^\mu}\bigg\rbrace_{\!\!q=0}.
\label{I10}
\ee
Then, setting $r=-k-q$ and $p=k$ in the general decomposition of \1eq{tensB}, we find 
\begin{align}
 \bigg\lbrace \frac{\partial {B}_{\sigma\rho}}{\partial q^\mu}\bigg\rbrace_{\!\!q=0} &=2 B_1'  k_\mu g_{\sigma\rho} +2 (B_2'+B_3'-B_4'-B_5') k_\mu k_\sigma k_\rho+ {\overline B}_3 (k_\rho g_{\mu\sigma} + k_\sigma g_{\mu\rho}),
	\label{Bexpansion}
\end{align}  
where the prime denotes differentiation with respect to $(k+q)^2$ and subsequently taking the limit $q\to 0$,
\be
	f'(-k,k) \equiv \, \lim_{q\to 0} \left\{ \frac{\partial f(q,-k-q,k)}{\partial\, (k+q)^2} \right\}, 
\label{Der}
\ee
and we have set \mbox{${\overline B}_3 \equiv B_3(0,k,-k)$}.

Our final step will be to substitute the r.h.s. of~\1eq{Bexpansion}  into \1eq{I10}, and make use of the elementary identities 
\begin{align}
	k^{\mu} \Gamma_{\mu\gamma\delta}^{(0)}(0,k,-k) &= 2k^2 P_{\gamma\delta}(k),\nonumber\\
	k^{\gamma} \Gamma_{\mu\gamma\delta}^{(0)}(0,k,-k) &= -k^2 P_{\mu\delta}(k), \nonumber\\
	k^{\delta} \Gamma_{\mu\gamma\delta}^{(0)}(0,k,-k) &= -k^2 P_{\gamma\mu}(k).
\label{elwis}
\end{align}
Evidently, the term proportional to  $B_1$ activates the first identity in~\1eq{elwis}, and therefore the longitudinal terms $\xi k^{\gamma} k^{\rho}/k^4$ and $\xi k^{\delta} k^{\sigma}/k^4$ contained in $\Delta^{\gamma\rho}(k)$ and $\Delta^{\delta\sigma}(k)$, respectively, vanish. In addition, the term in \1eq{Bexpansion} proportional to  $k_{\mu} k_{\sigma} k_{\rho}$ vanishes, again due to the fact that the first identity \1eq{elwis} is triggered by $k_{\mu}$, while \mbox{$k_{\sigma} k_{\rho} \Delta^{\gamma\rho}(k) \Delta^{\delta\sigma}(k) = \xi^2 k_{\gamma} k_{\delta}/k^4$}. The contribution of the term proportional to ${\overline B}_3$ may be obtained following similar considerations, employing the second and third identities in~\1eq{elwis}. Thus, one finally reaches the expression  
\be
I_{1}(0)=  i \frac{(d-1)}{d} C_A  \int_k \Delta(k^2) \left[\xi {\overline B}_3(k^2)-2 k^2\Delta(k^2) B'_1(k^2)\right].
\label{I1f}
\ee

In order to further evaluate the r.h.s of \1eq{I1f}, additional information on the behaviour of the form factors $B'_1(k^2)$ and  ${\overline B}_3(k^2)$ is required. As we will see in the next section, the BSE that controls the dynamics of the vertex function $B_{\alpha\beta}$ will furnish the approximate form of $B'_1(k^2)$, whereas, in the simplifying scheme that we will employ,  ${\overline B}_3(k^2)$ will be simply neglected throughout.

\section{\label{sec:bse}The BSE of the massless bound-state poles for general $\xi$} 

\begin{figure}[!t]
\includegraphics[scale=0.625]{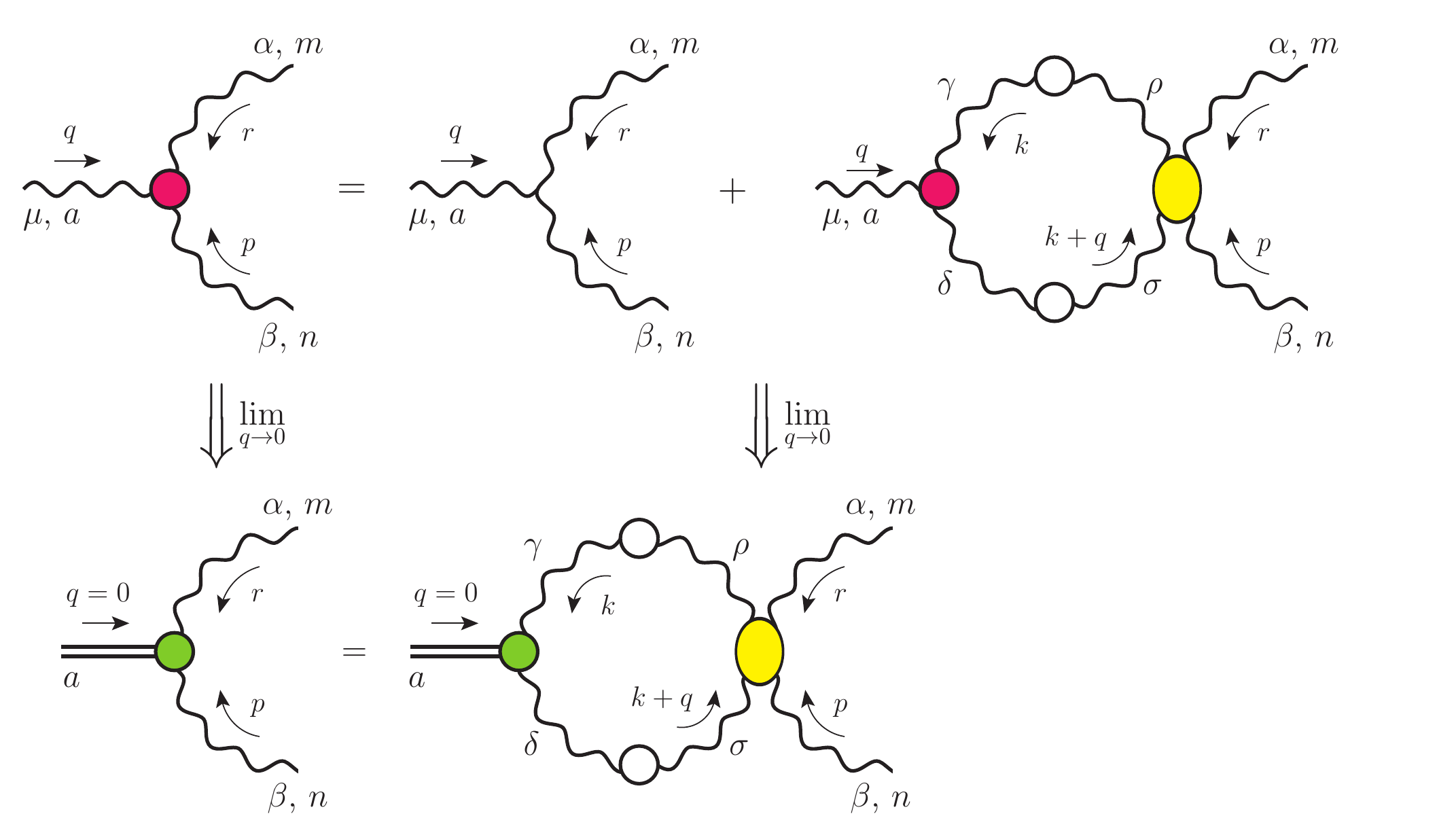}
\caption{\label{bse}The BSE for the bound-state wave function $B_{\alpha\beta}$ (lower line) is obtained by taking the $q\to0$ limit of the conventional BSE satisfied by the full vertex $\Gamma$ (upper line).}
\end{figure}

The BSE that determines $B_{\alpha\beta}$ may be derived following the procedure first outlined in~\cite{Jackiw:1973ha}, and more recently in~\cite{Aguilar:2011xe}. Specifically, the main steps may be summarized as follows.

\begin{enumerate}
\item We begin with the SDE for the three-gluon vertex, given pictorially in the first line of~\fig{Gammaprime}, and switch to the BS version of the same equation (first line of~\fig{bse}), by replacing the tree-level three-gluon vertex $\Gamma^{(0)}$ inside diagram ($a_1$) by a fully dressed one, $\Gamma$, and, correspondingly, the SD kernel by the BS kernel\footnote{The BS and SD kernels are different, because certain classes of diagrams, such as ladder graphs, which are allotted to the ``dressing'' of the vertex, must be excluded from the BS kernel, in order to avoid overcounting. For the general non-linear integral equation relating the two kernels see, {\it e.g.},~\cite{Bjorken:1965zz} and~\cite{Roberts:1994dr}.}. 

\item The next step is to substitute the full $\Gamma$ appearing on both sides of the BSE by the r.h.s. of \1eq{npandp}, multiply by $q^2$ to eliminate the pole contained in $\gp$, and take the limit of both sides as $q \to 0$.  In doing so, the term $q^2 \gnp$ vanishes faster than $I_1(q^2) B_{\alpha\beta}$  [we have used~\1eq{UIB}]; indeed, up to possible IR logarithmic divergences~\mbox{\cite{Aguilar:2013vaa,Pelaez:2013cpa,Blum:2014gna,Eichmann:2014xya,Cyrol:2016tym,Athenodorou:2016oyh,Duarte:2016ieu}}, the former term vanishes as ${\cal O}(q^2)$, while the latter as ${\cal O}(q)$ [see~\1eq{Bexpansion}]. 

\item 
Then, noticing that the factor $I_1$ cancels from both sides of the BSE, one finally arrives at [second line of \fig{bse}]
\begin{equation}
\lim_{q\to 0} B_{\alpha\beta}^{amn}(q,r,p)
= \lim_{q\to 0} \left\{ \int_k B_{\gamma\delta}^{abc}(q,k,-k-q)\Delta^{\gamma\rho}(k)\Delta^{\delta\sigma}(k+q){\cal K}_{\rho\alpha\beta\sigma}^{bmnc}(-k,r,p,k+q)\right\}.
\label{BSEq}
\end{equation}

\end{enumerate}

Before entering into the algebraic details necessary for the further evaluation of this equation, it is useful to identify qualitatively the ways in which the deviations from the Landau gauge are bound to manifest themselves. Specifically, the BSE consists of three distinct parts: \n{i} A part that retains the exact same form as in the Landau gauge, but now the quantities involved depend implicitly on $\xi$; we will refer to such terms as {\it``Landau-like''}. \n{ii} In addition,  the explicit $\xi$-terms coming from the gluon propagators will introduce new contributions that may be classified into two types: \n{ii$_{\,a}$} contributions that still involve the same unknown  quantity as before; for example, the BSE in the Landau gauge involves only the form factor $B_1$, because all others are annihilated by the transverse projectors multiplying the gluon propagators. Now, that particular form factor gets multiplied by various powers of $\xi$ (and modified combinations of Green's functions). \n{ii$_{\,b}$} At the same time, due to the non-transversality of the gluon propagators, terms involving other components of $B_{\alpha\beta}$  (which are absent in the {\it``Landau-like''} part) make also their appearance. Given the considerable technical complexity of the problem at hand, in what follows we will restrict ourselves to the study of the contributions \n{i} and \n{ii$_{\,a}$} of the BSE, neglecting all contributions that are of the type \n{ii$_{\,b}$}.

\begin{figure}[!t]
\includegraphics[scale=0.625]{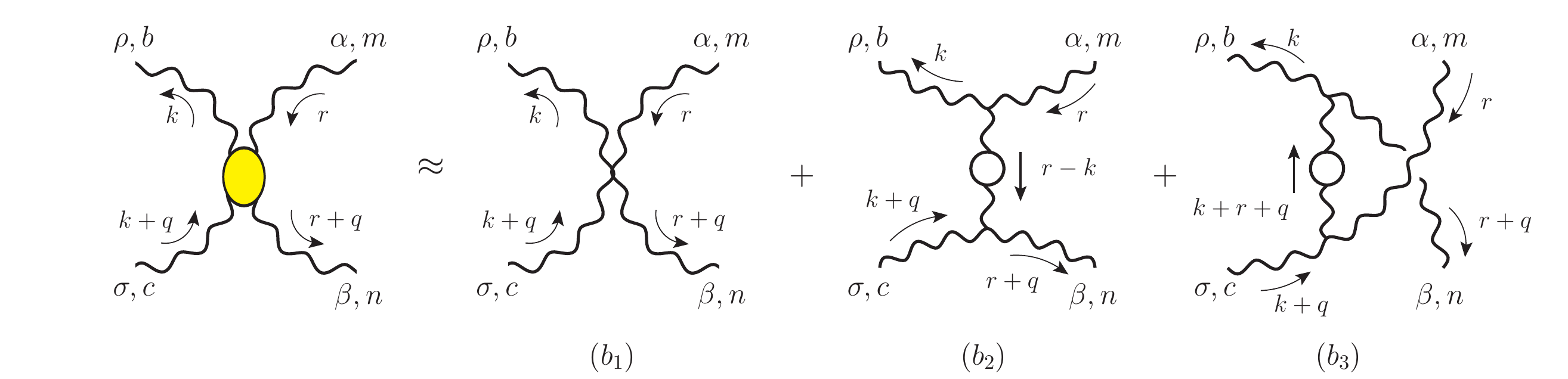}
\caption{\label{4gkernel}The lowest order approximation to the BSE kernel ${\cal K}_{\sigma\rho\nu\gamma}^{ncbm}$.}
\end{figure}

To proceed further, let us approximate the  four-gluon BS kernel ${\cal K}$ by its lowest-order set of diagrams shown in~\fig{4gkernel}. Then, the contribution to the BSE~\noeq{BSEq} due to the tree-level diagram $(b_1)$ is
\begin{align}
	(b_1)\to-\frac32iC_Ag^2f^{amn}\int_k\!B_{\gamma\delta}(q,k,-k-q)\Delta^{\gamma\rho}(k)\Delta^{\delta\sigma}(k+q)(g_{\alpha\rho}g_{\beta\sigma}-g_{\beta\rho}g_{\alpha\sigma}).
\label{b1diagram}	
\end{align}

For the one-loop dressed diagrams $(b_2)$ and $(b_3)$, which carry a statistical factor of $1/2$, we consider the internal propagators to be fully dressed, whereas for the three gluon vertex we use the Ansatz
\begin{align}
	\Gamma_{\mu\alpha\beta}(q,r,p)=f(r)\Gamma^{(0)}_{\mu\alpha\beta}(q,r,p),
	\label{vertf}
\end{align} 
{\it i.e.}, we multiply the three-level expression by a (possibly $\xi$-dependent) function of a single kinematic variable, which is used to model in a tractable way some of the possible nonperturbative corrections associated with the full vertex. These two diagrams contribute to~\noeq{BSEq} the term
\begin{align}
	(b_2)&+(b_3)\to  \frac14g^2C_Af^{amn}\int_k\!B_{\gamma\delta}(q,k,-k-q)\Delta^{\gamma\rho}(k)\Delta^{\delta\sigma}(k+q) \\
	&\times\left[f^2(k-r)\Gamma^{(0)}_{\rho\mu\alpha}(-k,k-r,r)\Gamma^{(0)}_{\sigma\nu\beta}(k+q,r-k,-r-q)\Delta^{\mu\nu}(r-k)\right. \nonumber \\
	&\left.-f^2(k+q+r)\Gamma^{(0)}_{\rho\mu\beta}(-k,k+q+r,-r-q)\Gamma^{(0)}_{\sigma\nu\alpha}(k+q,-r-k-q,r)\Delta^{\mu\nu}(k+r+q)\right]. \nonumber
\label{b23diagram}	
\end{align} 

Next, we expand both sides of~\noeq{BSEq} to leading order in $q$, and take the limit $q \to 0$; in addition, we isolate the $B'_1$ contribution by contracting with $P^{\alpha\beta}(r)$, and finally neglect on the r.h.s. all form factors and their derivatives except for $B_1'$; this last step eliminates all contributions of the type \n{ii$_{\,b}$}, as announced above. After implementing these operations, one arrives at the result
\begin{align}
	3(q\!\cdot\! r)B'_1(r^2)& = \frac{g^2}4C_A\int_k(q\cdot k)B'_1(k^2) \Delta^{\rho}_{\delta}(k)\Delta^{\delta\sigma}(k)P^{\alpha\beta}(r)\Big[-6i(g_{\alpha\rho}g_{\beta\sigma}-g_{\beta\rho}g_{\alpha\sigma})\nonumber \\
	&+f^2(k-r)\Gamma^{(0)}_{\rho\mu\alpha}(-k,k-r,r)\Gamma^{(0)}_{\sigma\nu\beta}(k,r-k,-r)\Delta^{\mu\nu}(r-k) \nonumber \\
	&-f^2(k+r)\Gamma^{(0)}_{\rho\mu\beta}(-k,k+r,-r)\Gamma^{(0)}_{\sigma\nu\alpha}(k,-r-k,r)\Delta^{\mu\nu}(k+r)\Big].
\end{align}
Finally, we notice that the first term in square brackets, coming from the tree-level diagram $(b_1)$, does not contribute to the BSE for any value of $\xi$, whereas the simple change in the integration variable ($k\to -k$) reveals that the last two terms are equal and add up, to finally give
\begin{align}
	3(q\!\cdot\! r)B'_1(r^2)& =
	-\frac{g^2}2C_Aq^\lambda\int_k B'_1(k^2)k_\lambda \Delta^{\rho}_{\delta} (k)\Delta^{\delta\sigma}(k)\Delta^{\mu\nu}(k+r)P^{\alpha\beta}(r)\nonumber \\
	&\times f^2(k+r)\Gamma^{(0)}_{\rho\mu\beta}(-k,k+r,-r)\Gamma^{(0)}_{\sigma\nu\alpha}(k,-r-k,r)\Delta^{\mu\nu}(k+r).
\end{align} 
Evidently, the r.h.s. integral displays only one free Lorentz index and the unique momentum scale $r$, so that it can only be proportional to $r_\lambda$; thus, the scalar product $q\!\cdot\! r$ drops out from both sides, and one is left with the final equation
\begin{align}
	B'_1(r^2)& = 
	-\frac{2\pi}3\alpha_s C_A\int_k\frac{r\!\cdot\!k}{r^2}  B'_1(k^2) \Delta^{\rho}_{\delta}(k)\Delta^{\delta\sigma}(k)\Delta^{\mu\nu}(k+r)P^{\alpha\beta}(r)\nonumber \\
	&\times f^2(k+r)\Gamma^{(0)}_{\rho\mu\beta}(-k,k+r,-r)\Gamma^{(0)}_{\sigma\nu\alpha}(k,-r-k,r)\Delta^{\mu\nu}(k+r)\nonumber \\
	&=\frac{2\pi}3\alpha_s C_A\int_kB'_1(k^2)\sum_{n=0}^3\xi^n{\cal A}_n(r,k),
	\label{bsemink}
\end{align} 
where $\alpha_s = g^2/4\pi$, and 
\begin{align}
{\cal A}_0(r,k) &= \frac{4(r\!\cdot\!k)[r^2k^2-(r\!\cdot\! k)^2]}{p^2k^2r^2(k-p)^2}[8p^2k^2 + 6(r\!\cdot\!k)(r^2+k^2)+3(r^4+k^4)+(r\!\cdot\! k)^2]\nonumber \\
&\times f^2(k+r)\Delta^2(k)\Delta(k+r), 
\nonumber\\
{\cal A}_1(r,k) &= \frac{(r\!\cdot\!k)(k^2-r^2)^2}{r^2(k+r)^4}\left[2 + \frac{(r\!\cdot\! k)^2}{k^2p^2} \right]f^2(k+r)\Delta^2(k),
\nonumber\\
{\cal A}_2(r,k) &= \frac{(r\!\cdot\!k)[(k+r)^2-r^2]^2}{k^6r^2} \left[3 - \frac{r^2k^2-(r\!\cdot\! k)^2}{(k+r)^2p^2} \right] f^2(k+r)\Delta(k+r),
\nonumber\\
{\cal A}_3(r,k) &= \frac{(r\!\cdot\!k)[r^2k^2-(r\!\cdot\! k)^2]}{k^6 (k+r)^4}f^2(k+r). 
\label{A0123}
\end{align}
 
Passing~\1eq{bsemink} to Euclidean space, where $\int \d^4k \to i\int \d^4k_{\s E}$,  requires use of the standard transformation rules $(k^2, r^2, k\!\cdot\! r) \to - (k_{\s E}^2, r_{\s E}^2, \,k_{\s E}\!\cdot\! r_{\s E})$, with $k_{\s E}^2,  r_{\s E}^2 \ge 0$,  together with \mbox{$\Delta(-k^2_\mathrm{\s E}) \to - \Delta_\mathrm{\s E}(k^2_\mathrm{\s E})$} and $B'_1(-k^2_\mathrm{\s E}) \to - B'_{1\s{\mathrm{E}}}(k^2_\mathrm{\s E})$. Dropping the subscript `E' throughout, writing the integration measure in spherical coordinates,
\begin{equation}
\int\frac{\d^4k_{\s E}}{(2\pi)^4} = \frac{1}{(2\pi)^3}\int_0^\infty \d y\,y\int_0^\pi \d\theta\, \sin^2\theta,
\end{equation}
and setting
\begin{align}
	x&\equiv r^2;& \quad y&\equiv k^2;&
	z&\equiv (r+k)^2 = x+y + 2 \sqrt{xy}\cos\theta,
\end{align}
the BSE of \1eq{bsemink} becomes then
\bea
B'_1(x) &=& -\frac{\alpha_s C_A}{12\pi^2}\int_0^\infty\! \d y\, B'_1(y) \int_0^\pi \!\d\theta\,\sin^2\theta\cos\theta\sum_{n=0}\xi^n A_n(x,y,\theta),
\label{euclideanBS}
\eea
with\footnote{Note that in writing $A_2$ we have used that $(z-x)^2 = y^2 \left(1+ \sqrt{\frac{x}{y}} \cos\theta\right)^2$ in order to cancel a factor of $y^2$ that appeared in the denominator; this final form of $A_2$ turns out to be more appropriate for the numerical treatment of the resulting equation.}
\begin{align}
A_0 &= y\sin^2\theta\left[z+10(x+y)+\frac{1}{z}(x^2+y^2+10xy)\right]\sqrt{\frac{y}{x}}f^2(z)\Delta^2(y)\Delta(z),
\nonumber\\
A_1 &=  y(y-x)^2 (2+\cos^2\theta)\sqrt{\frac{y}{x}}f^2(z) \frac{\Delta^2(y)}{z^2},
\nonumber\\
A_2 &=  y\left(1+2\sqrt{\frac{x}{y}}\cos\theta\right)^2 (3z - y\sin^2\theta)\sqrt{\frac{y}{x}} f^2(z)\frac{\Delta(z)}{zy}, 
\nonumber\\
A_3 &=   \frac{x^2}{y z^2}\sqrt{\frac{y}{x}}f^2(z)\sin^2\theta.
\label{theAs}
\end{align}

Before proceeding to the numerical treatment of the equation found, let us study its IR limit. To that end, we carry out the Taylor expansion of the  
$A_i$ terms around $x=0$, and then perform the angular integration; this yields the results
\begin{align}
	\int_0^\pi \!\d\theta\,\sin^2\theta\cos\theta\, A_0(x,y,\theta)&\approx \frac32 \pi y^3 f(y) \Delta(
  y)^2 [2 \Delta(y) f'(y) + 
   f(y) \Delta'(y)],\nonumber \\
   \int_0^\pi \!\d\theta\,\sin^2\theta\cos\theta\, A_1(x,y,\theta)&\approx \frac54\pi y\Delta^2(y)[yf'(y)-f(y)],\nonumber \\
   \int_0^\pi \!\d\theta\,\sin^2\theta\cos\theta\, A_2(x,y,\theta)&\approx\frac\pi8[5y(2\Delta(y)f'(y)+\Delta'(y)f(y))+11f(y)\Delta(y)],\nonumber \\
    \int_0^\pi \!\d\theta\,\sin^2\theta\cos\theta\, A_3(x,y,\theta)&\approx\frac\pi4\frac{x^2}{y^3}f(y)[yf'(y)-f(y)].
\end{align}
   
Suppose now that we are interested only in solutions of the BSE~\noeq{euclideanBS} that are well behaved in the IR\footnote{In principle, an IR divergent $B'_1$ could still lead to a finite value for the integral~\noeq{I1f}; we will nevertheless exclude this case from our analysis.}. Then, we see that the non-negative $y$ powers appearing in the first three terms render them IR stable, provided that the form factor $f$ displays at most a log-like divergence (recall that for all $\xi$ studied the gluon propagator saturates in the IR). Indeed, numerically these terms drive the solution to acquire a positive IR value, $B'_1(0)>0$. On the other hand, 
the term $A_3$ is far more restrictive; IR finite solutions can be only achieved for $f$ such that
\begin{align}
	f(y)[yf'(y)-f(y)]&=c;& f(y)&=\pm\sqrt{c+(c_1y)^2},
	\label{potdiv}
\end{align}
where $c>0$, and $c_1$ is an integration constant. Thus, we see that, within the approximations employed, 
the presence of the term $A_3$ makes the BSE incompatible with an IR divergent~$f$.

\begin{table}[!t]
\begin{center}
\begin{tabular}{ccccc}
\hline\hline
{$\xi$}&\hspace{1.7cm} $g^2_1$ \hspace{.85cm} & \hspace{.85cm}  $\rho_1$\hspace{.85cm} & \hspace{.85cm} $\rho_2$\hspace{.85cm} & \hspace{.475cm} $m_0$ [MeV] \\
\hline
0 & 5.60 & 26.63 & 1.96 & \hspace{.475cm}  391 \\
0.1 & 5.70 & 24.16 & 2.03 & \hspace{.475cm} 404\\
0.2 & 5.73 & 22.05 & 2.03 & \hspace{.475cm} 411\\
0.3 & 5.86 & 19.89 & 2.16 & \hspace{.475cm} 426\\
0.4 & 5.91 & 18.54 & 2.10 & \hspace{.475cm} 427\\
0.5 & 6.01 & 19.67 & 1.99 & \hspace{.475cm} 417\\
\hline
\end{tabular}
\end{center}
\caption{\label{params}Fitting parameters employed in~\1eq{fit_gluon}.}
\end{table}

\section{\label{sec:numan} Numerical analysis}

In this section we carry out a detailed numerical analysis of the BSE in~(\ref{euclideanBS}), which controls the formation of the massless bound-state.

In order to solve  numerically this homogeneous integral equation, it is convenient to  convert it to  an eigenvalue problem; in particular, one writes   
\begin{align}
	B_1^{\prime}(x) = \lambda \int \, \widetilde{\mathcal K}(x,y,\theta) B_1^{\prime}(y),
\label{eigenproblem}
\end{align}
where the extra parameter $\lambda$ acts as an eigenvalue, $B_1^{\prime}(x)$ represents the corresponding eigenvector, and the kernel $\widetilde{\mathcal K}$, together with the corresponding integration measure, may be straightforwardly deduced from~\1eq{euclideanBS}. Evidently, in order to recover the original integral equation, we are looking for nontrivial solutions corresponding to the eigenvalue $\lambda=1$. However, notice that, due to the homogeneity of the BSE, if a solution $B'_1(x)$ exists, then it is clear that the family of function $cB'_1(x)$, where $c$ is any real constant, are also equally acceptable solutions. The way to resolve this ambiguity is by supplementing the BSE with a judiciously chosen boundary condition. 

\begin{figure}[!t]
\includegraphics[scale=0.61]{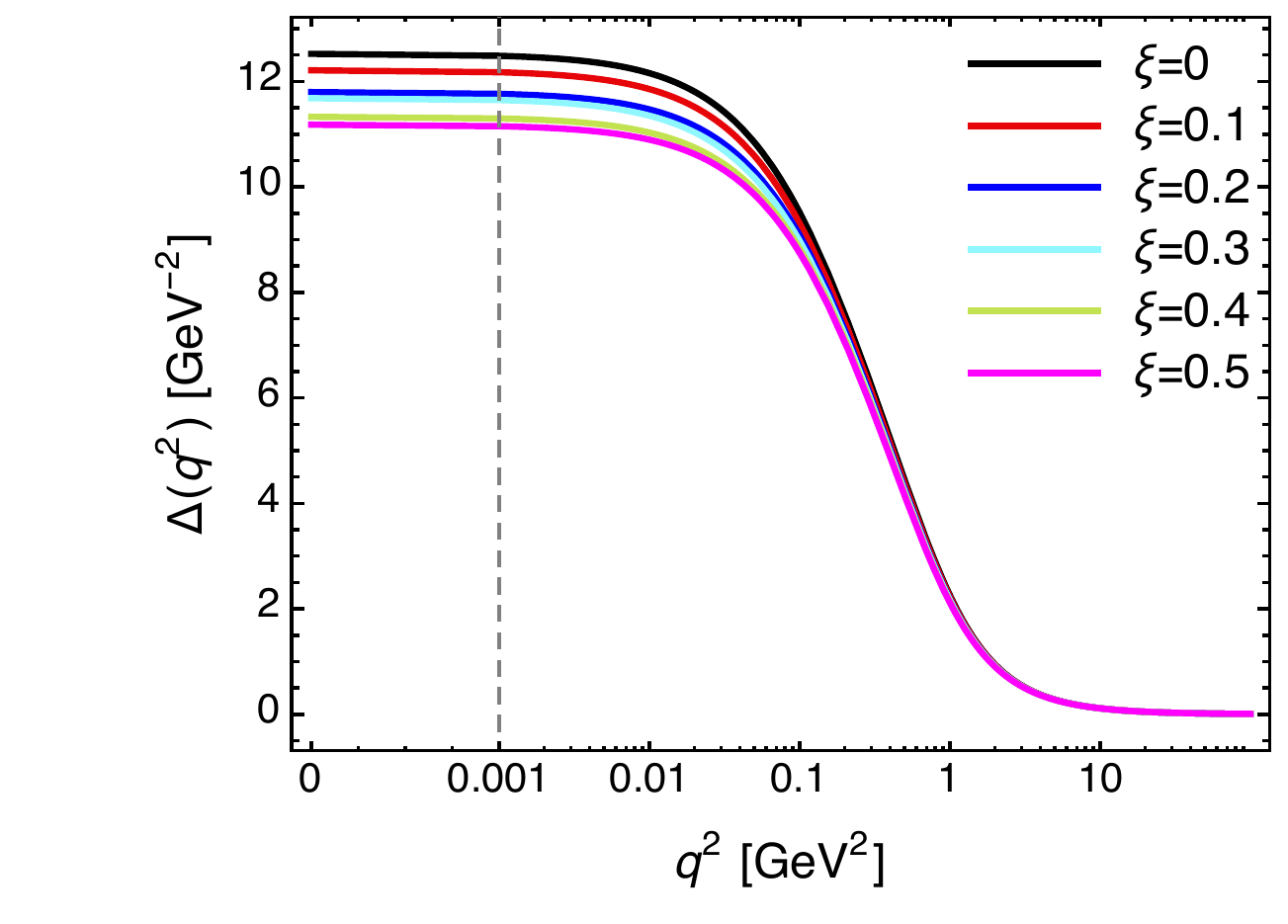}
\caption{\label{Deltas}The gluon propagator for $\xi=0$  through $\xi=0.5$ obtained from the fit~\noeq{fit_gluon} to the lattice data of Ref.~\cite{Bicudo:2015rma} using the parameters reported in Table~\ref{params}. In order to expose the saturation value at zero, the momentum scale in the abscissa is turned from logarithmic to linear after the vertical dashed line.}
\end{figure}

To understand the origin of the boundary condition that we will use, let us first point out that the gluon propagators composing the kernel of~\noeq{euclideanBS} will be treated as {\it external} objects, in the sense that we will use for them a fit to the data obtained from the lattice simulation of~\cite{Bicudo:2015rma}, 
corresponding to the gluon propagators $\Delta(q^2,\xi)$ with $\xi\in[0,0.5]$. As discussed in~\cite{Bicudo:2015rma}, the IR finiteness of the gluon propagator persists also in the $\xi\neq0$ cases; thus, a physically motivated fit describing the data is given by (see~\fig{Deltas})
\begin{align}
	\Delta^{-1}(q^2,\xi) &= m^2(q^2) + q^2\left[1+\frac{C_{\rm A}}{32\pi^2} \left(\frac{13}{3} - \xi\right)g_1^2\ln\left(\frac{q^2+ \rho_1 m^2(q^2)}{\mu^2}\right)\right];\nonumber \\
	m(q^2) &= \frac{m_0^4}{q^2 + \rho_2 m_0^2}, 
	\label{fit_gluon}
\end{align}   
with $g^2_1$, $\rho_1$, $\rho_2$ and $m_0$ fitting parameters, whose best values are reported in Table~\ref{params}. 

Let us now assume that a solution $B'_1(x)$ has been found (with $\lambda=1$) using a given $\Delta(q^2,\xi)$ as input into \1eq{euclideanBS}, together with an Ansatz for the function $f(z)$ (see discussion below); then, within the approximations employed, the constant $c$ is uniquely determined by demanding that 
the ``input'' and ``output'' values for $\Delta^{-1}(0)$ {\it coincide}, namely 
\begin{align}
	c&=\sqrt{\Delta^{-1}(0;\xi)/I_1(0)},
\end{align}
where $I_1(0)$ is the four-dimensional Euclidean version of~\1eq{I1f},
\begin{align}
	I_1(0) &=\frac{3 C_A}{32\pi^2} \int_0^\infty\! \mathrm{d}y\,[y\Delta(y)]^2B'_1(y)\,. 
\end{align}

\begin{figure}[!t]
\hspace{-0.5cm}\includegraphics[scale=0.61]{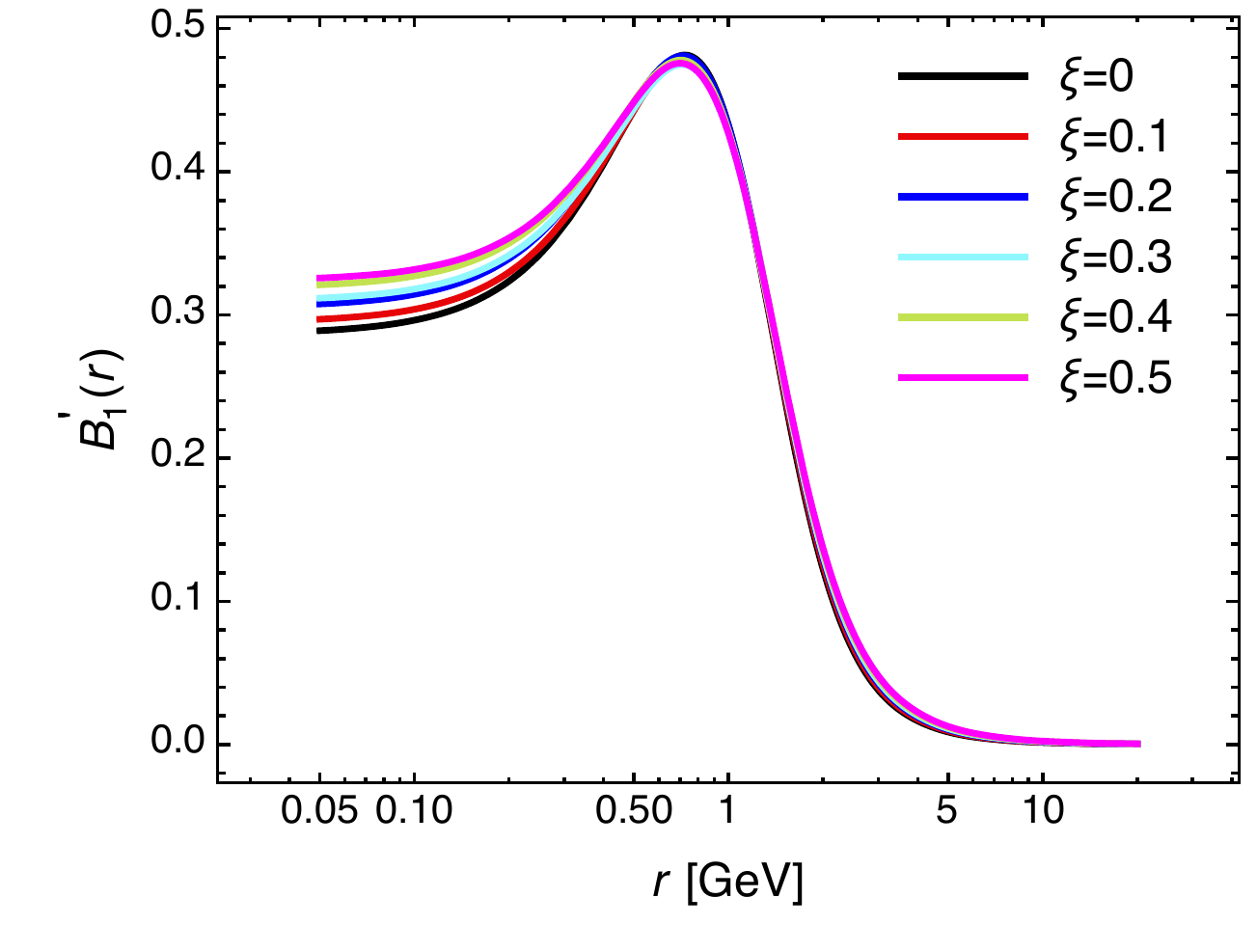}\hspace{.5cm}
\includegraphics[scale=0.61]{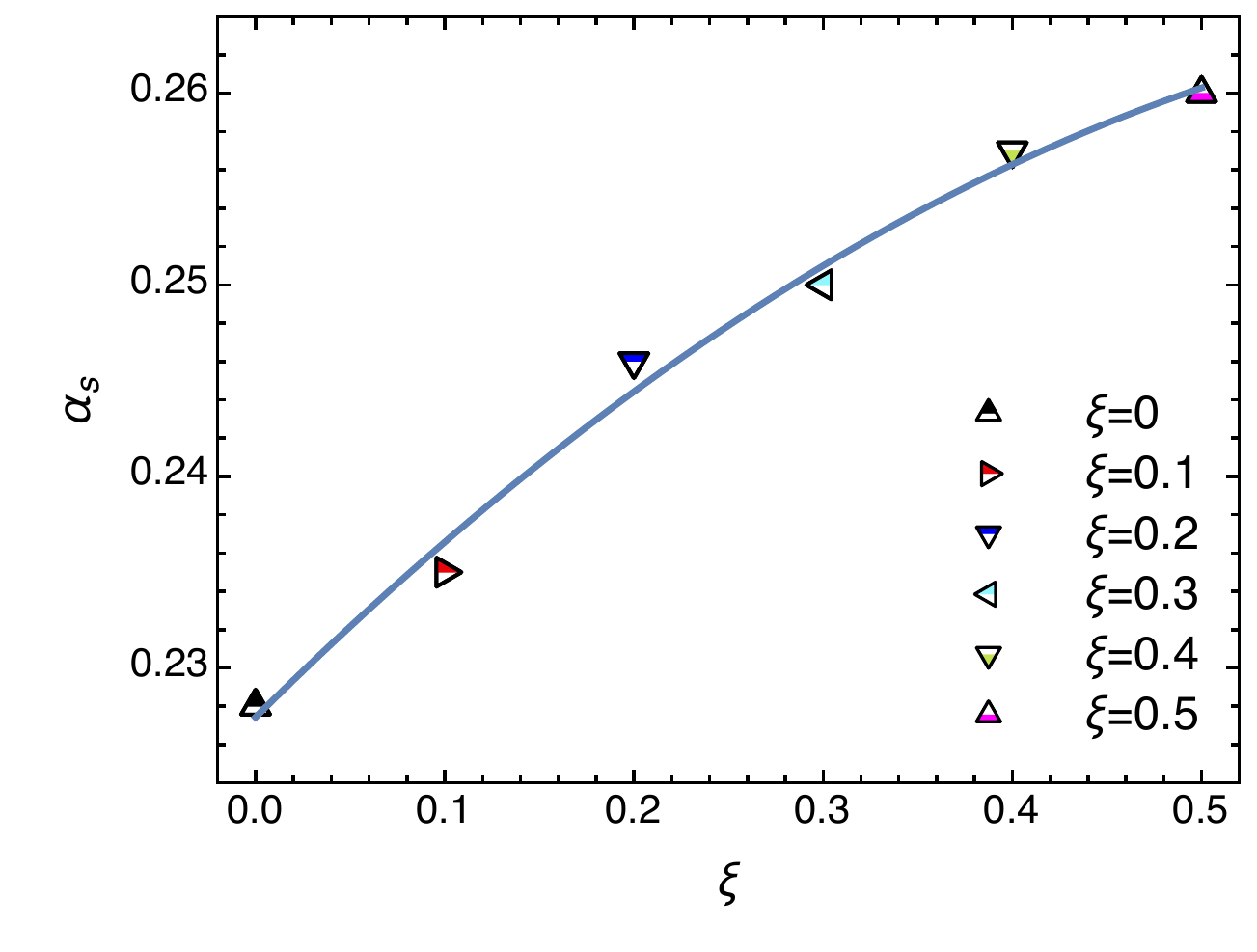}
\caption{\label{fig:bse_tree}Solution of the BS equation~\noeq{eigenproblem} obtained in the tree-level case $f(z)=1$ (left), and the corresponding value of the strong coupling $\alpha_s$ evaluated at the renormalization scale $\mu=4.3$ [GeV] (right). Notice the quadratic dependence of the coupling on the gauge-fixing parameter $\xi$, with $\alpha_s(\xi)=0.227+ 0.098\xi-0.064 \xi^2$ (solid line in the right panel).}
\end{figure}

The next issue to address is the role of the vertex form factor $f(z)$, which enters (quadratically) in the kernel of~\1eq{euclideanBS} through~\1eq{vertf}. Given that the main purpose of introducing $f(z)$ is to account for deviations of the three gluon vertex from its tree-level value, it is natural to first inquire what happens if we  set $f(z)=1$, \ie assume that $\Gamma = \Gamma^{(0)}$.   In order to set the stage, let us point out that the strong charge $\alpha_s$ appearing in $\widetilde{\mathcal K}$ is defined in the momentum subtraction (MOM) scheme, which has been used for the renormalization of the propagators $\Delta(q^2,\xi)$. Within this scheme, and for $\mu=4.3$ GeV, the value of the strong charge has been estimated to be $\alpha_s =0.22$; the determination of this value entails a subtle combination of 4-loop perturbative results, nonperturbative information included in the vacuum condensate of dimension two, and the extraction of $\Lambda_{\rm QCD}$ from lattice results of the ghost-gluon vertex in the Taylor kinematics~\cite{Boucaud:2008gn}.

Then, with $f(z)=1$, our numerical analysis reveals that, in order to obtain nontrivial solutions (with $\lambda=1$), one has to choose a different value of $\alpha_s$ for each value of  $\xi$; in fact, $\alpha_s$ increases as $\xi$ is varied from $0$ to $0.5$, ranging from $0.23$ to $0.26$ (see \fig{fig:bse_tree}, right panel). The corresponding solutions of the BSE are shown in the left panel of~\fig{fig:bse_tree}. This departure of $\alpha_s$ from the MOM value of $0.22$ quoted above may be considered more than acceptable given the approximate nature of the BSE studied, as well as the theoretical uncertainties entering in the analysis of~\cite{Boucaud:2008gn}. Moreover, a nontrivial dependence of $\alpha_s$ on  $\xi$ is expected on theoretical grounds, given that the MOM $\beta$ function depends explicitly on  $\xi$ (see, \eg~\cite{Chetyrkin:2000dq}, and references therein).

Let us now turn to the main thrust of our analysis, and  allow $f(z)$ to vary.  
Specifically, we will fix  $\alpha_s$ to a given value (\eg at $0.22$) for {\it all} $\xi$, 
and ascribe the  residual $\xi$-dependence to the vertex form factor $f(z)$. 
In doing so, and according to the discussion in sections \ref{int} and \ref{sec:bse},  
one must distinguish two qualitatively different cases, depending on the IR behavior of the $f(z)$ that is employed 
at the point of departure, namely in the Landau gauge. In particular, we will consider
\n{i} two IR finite Ans\"atze for $f(z)$, denoted by  $f_1(z)$ and $f_2(z)$, and 
\n{ii} one Ansatz denoted by $f_3(z)$, which diverges logarithmically in the IR.

\n{i} When $f(z)$ is IR finite, a-priori one does not expect dramatic changes 
in the type of solutions obtained from the BSE as $\xi$ is varied. 
In fact, the additional strength that one must supply to the kernel in order to fulfil the condition $\lambda=1$ as  $\xi$ 
increases (which, when $f(z)=1$, is accomplished by raising $\alpha_s$), originates from an appropriate variation of $f(z)$. 
That being the case, one would expect that the dependence of $f(z)$ on  $\xi$ will turn out 
to be relatively mild (at least for the range of  $\xi$ that we consider), since, 
qualitatively speaking,  one ends up distributing the square root of the 
excess strength over a function that ranges from zero to infinity. 
   
In order to explore the veracity of this expectation numerically, 
we fix indeed \mbox{$\alpha_s=0.22$}, and employ the two aforementioned Ans\"atze, $f_1(z)$ and $f_2(z)$, which 
display different behaviors both at low and intermediate momenta, but are both finite at the origin. 

\begin{figure}[!t]
\hspace{-0.5cm}\includegraphics[scale=0.61]{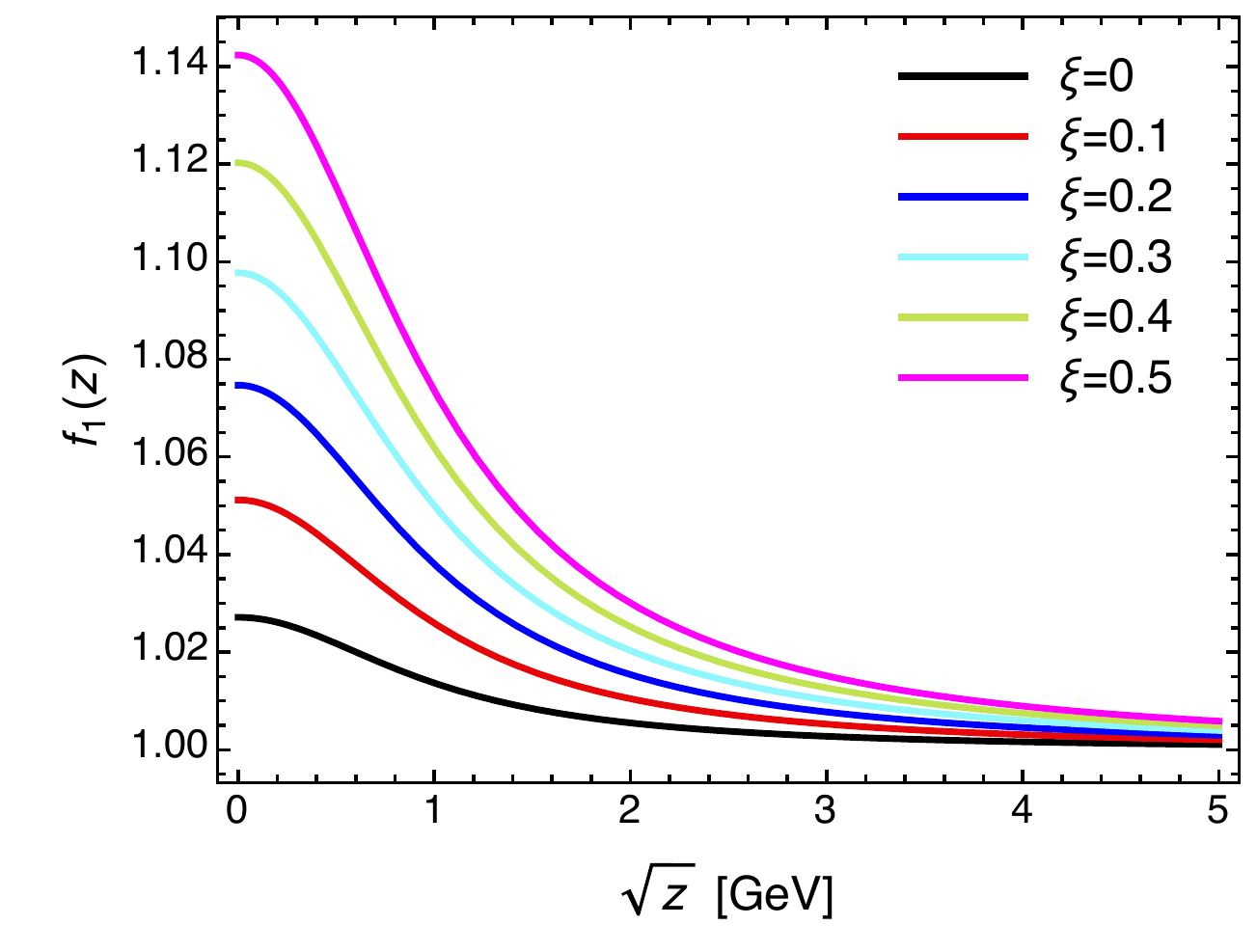}\hspace{.5cm}
\includegraphics[scale=0.61]{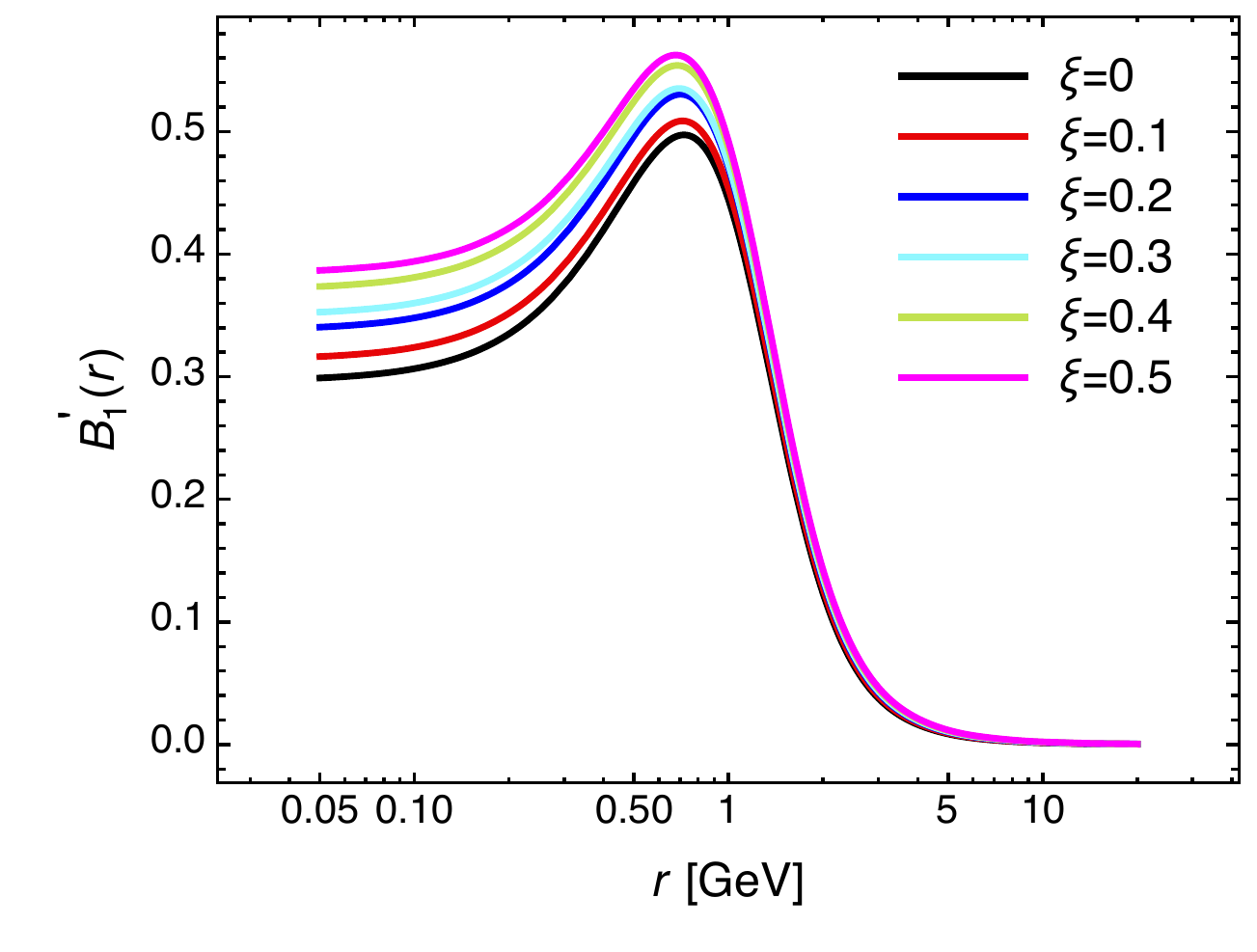}\\
\hspace{-0.5cm}\includegraphics[scale=0.61]{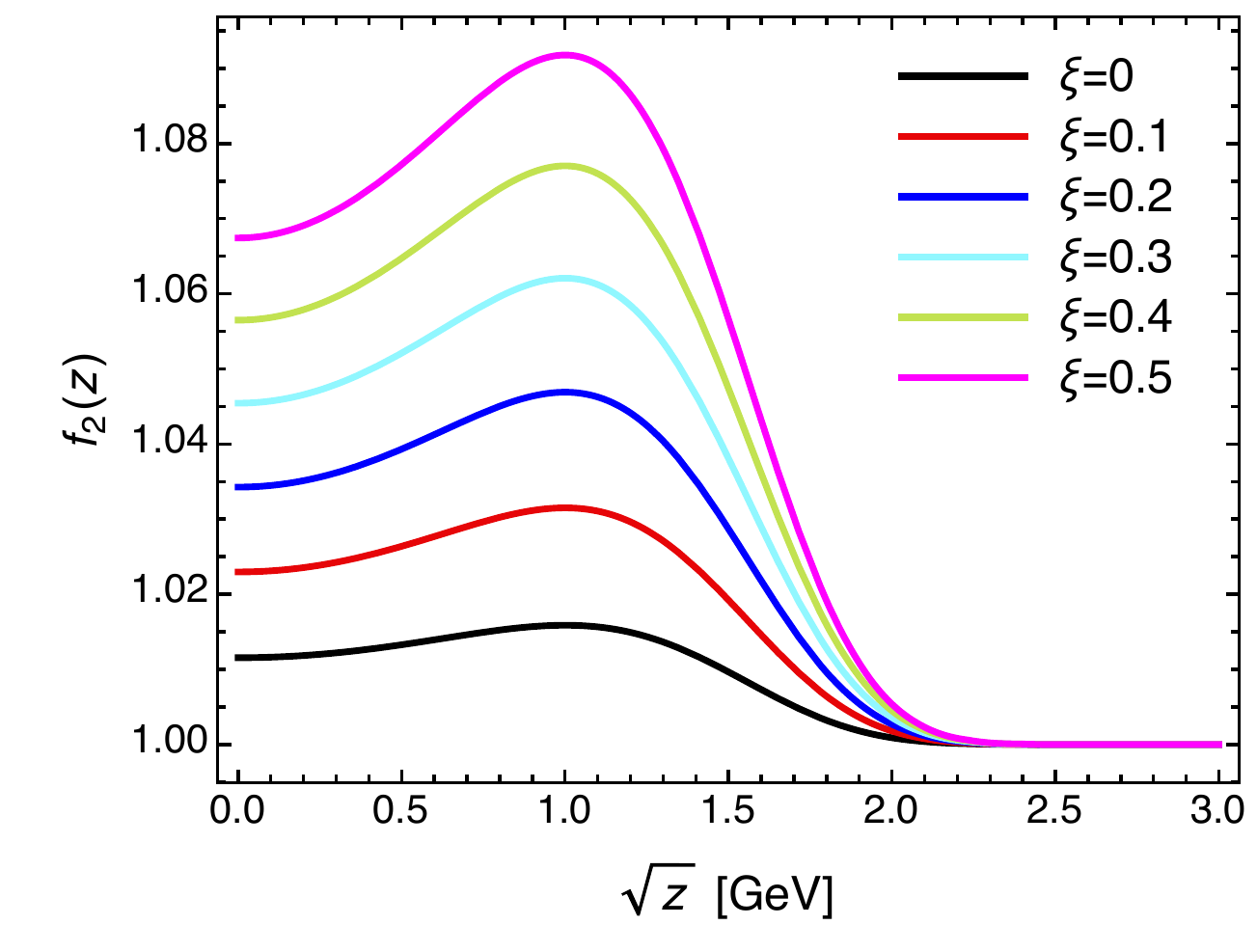}\hspace{.5cm}
\includegraphics[scale=0.61]{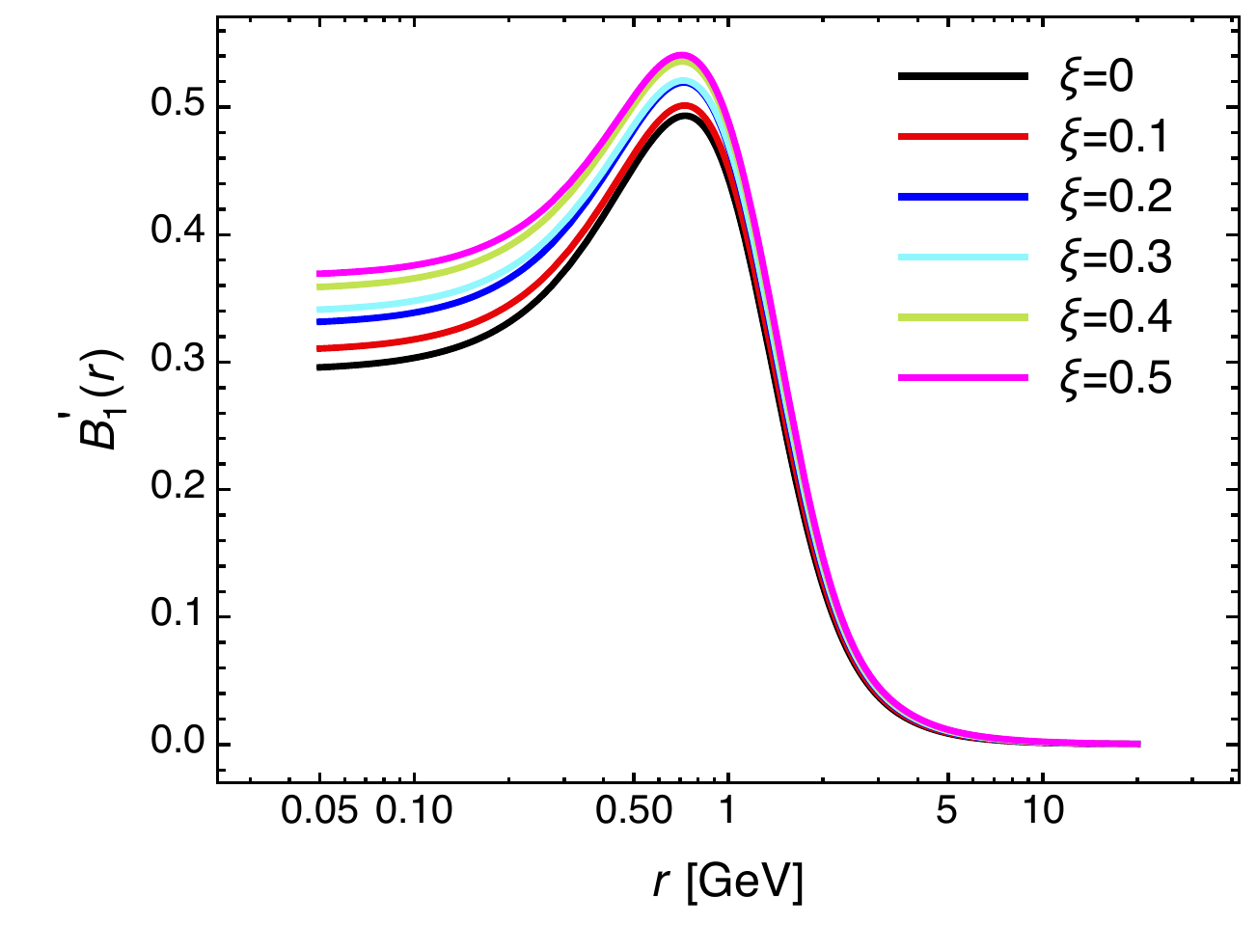}
\caption{\label{fig:bse_f12}The $\xi$ dependence of the form factors $f_1(z)$ (top left) and $f_2(z)$ (bottom left), and the corresponding solution of the BSE~\noeq{euclideanBS} obtained for $\alpha_s=0.22$ (top and bottom right).}
\end{figure}

Let us start with the simplest case, setting
\begin{align}
f_1(z) = \left[1 + \frac{(0.5\xi+a)}{1+z/z_0}\right]^{1/2},
\label{vertex1}
\end{align}
with $a=0.055$ and  $z_0 = 1$ GeV$^2$. On the top left panel of~\fig{fig:bse_f12} we show the shape of this form factor for different values of $\xi$. Clearly, we see that $f_1$ does not deviate considerably from unity: it has a maximum at $z=0$ (which increases for higher values of $\xi$), and tends monotonically to 1 in the ultraviolet region. The corresponding solutions of the BSE are shown on the top right panel of the same figure. One observes a mild dependence on the gauge fixing parameter: all $B'_1$ display a maximum in the region located around 1 GeV, whose  height increases for decreasing values of $\xi$. 

As a second model for the three-gluon vertex form factor let us consider the function
\begin{align}
	f_2(z) = \left\{1 + a(\xi+b)\exp\left[{\frac{-2(z-z_0)}{\omega^2}}\right]\right\}^{1/2},
\label{vertex2}
\end{align}
where $a = 0.32$ , $b = 0.1$, $z_0 = 1.0 \,\mbox{GeV}^2$ and $\omega = 2.5 \,\mbox{GeV}$. The resulting shapes for the different values of $\xi$ are shown on the left bottom panel of~\fig{fig:bse_f12}; one sees again the presence of a maximum and a nonvanishing value at $z=0$, the main difference with respect to $f_1$ being that now the behavior is not monotonic. The corresponding solutions of the BSE are shown on the bottom right panel of the same figure, where we see no qualitative (and almost no quantitative) change in their behavior with respect to the previous case. 

\begin{figure}[!t]
\hspace{-0.5cm}\includegraphics[scale=0.61]{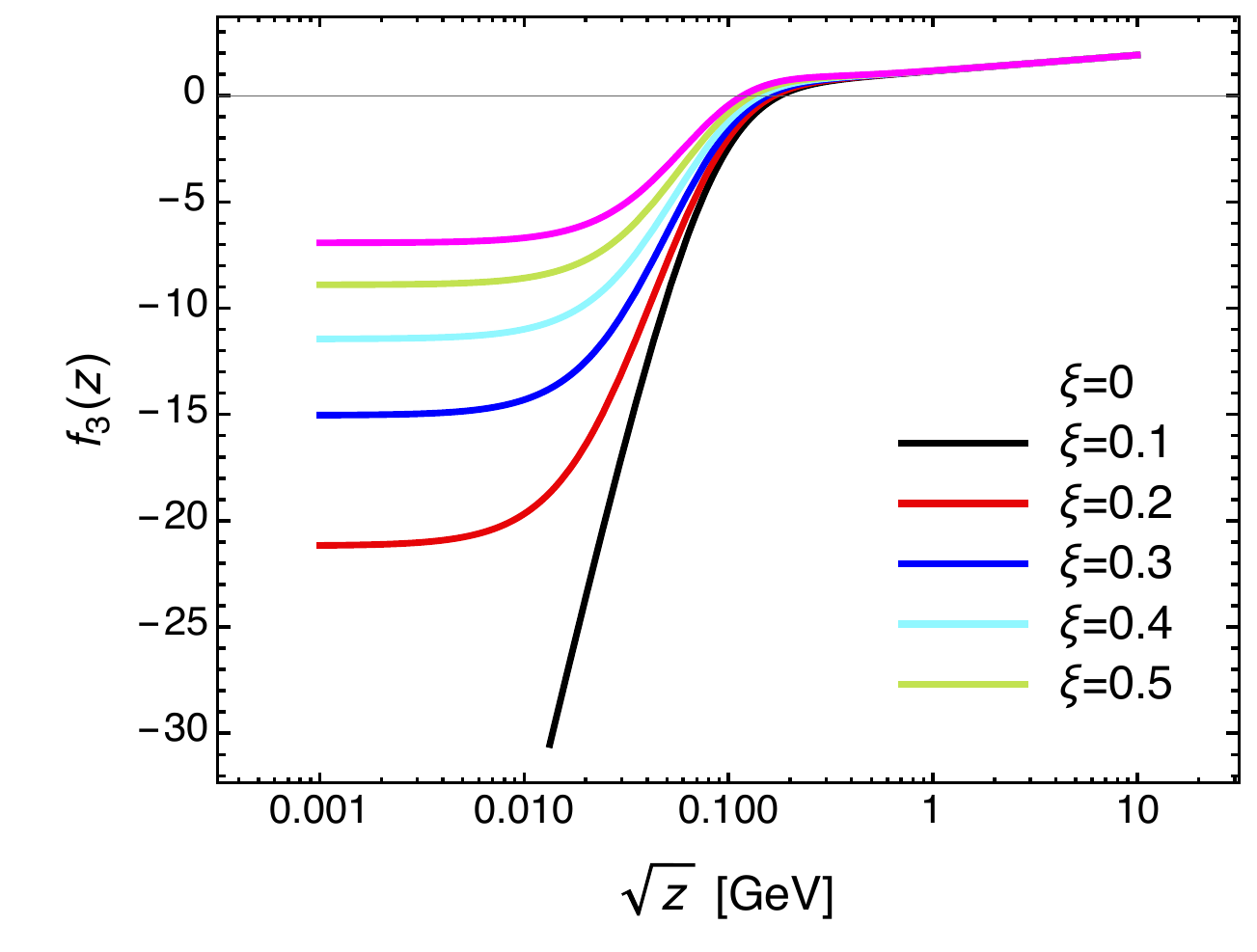}\hspace{.5cm}
\includegraphics[scale=0.61]{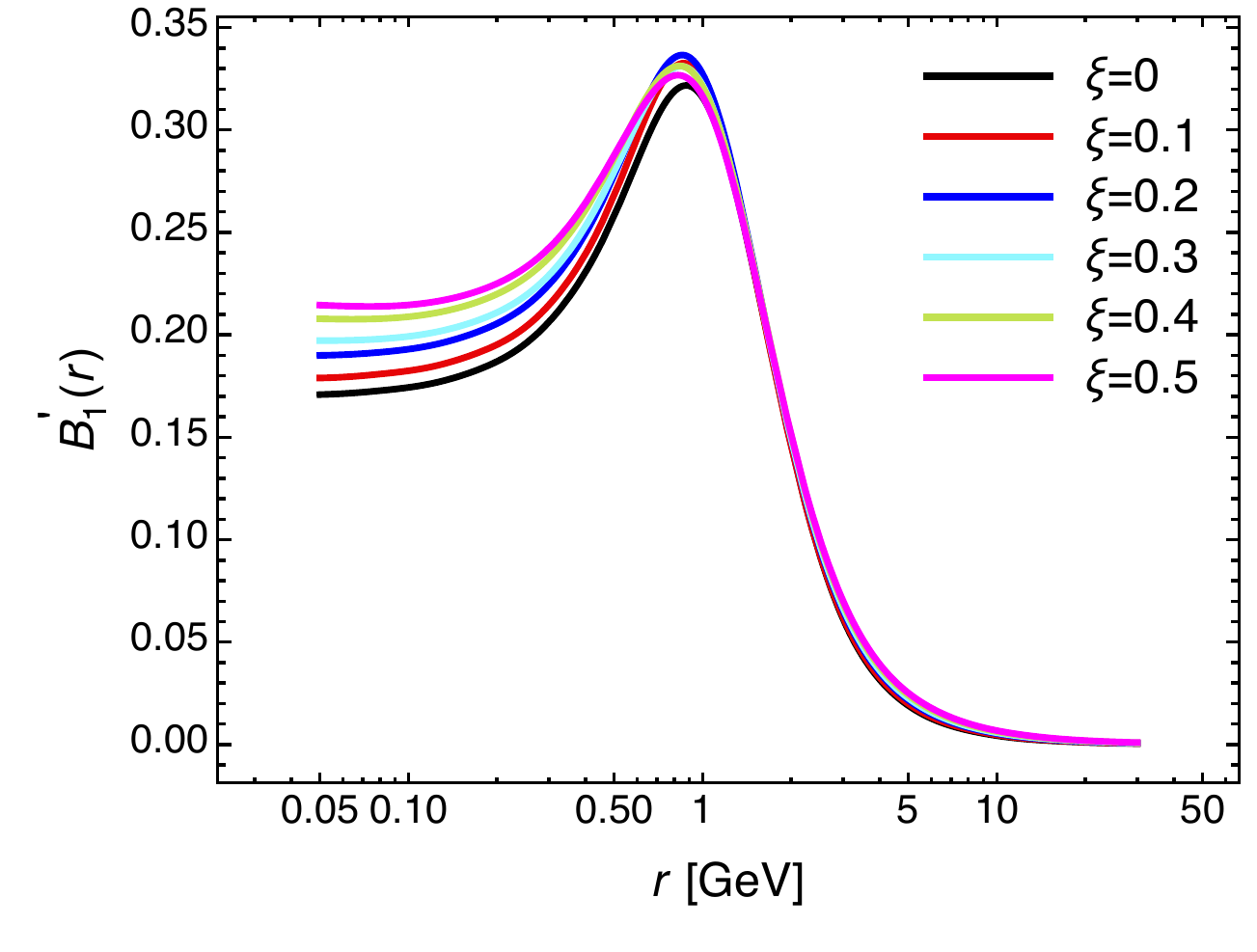}
\caption{\label{fig:bse_f3} The $\xi$ dependence for the three gluon vertex model, $f_3(z)$,  given in Eq.~(\ref{vertex3}) (left panel) and, the corresponding solution for the BS equation~(\ref{euclideanBS}) (right panel).}
\end{figure}

\n{ii}
The third and final form factor $f_3$ is motivated by recent nonperturbative studies of the three-gluon vertex in the Landau gauge~\cite{Aguilar:2013vaa,Pelaez:2013cpa,Blum:2014gna,Eichmann:2014xya,Cyrol:2016tym,Athenodorou:2016oyh,Duarte:2016ieu}. Specifically, in certain kinematic limits characterized by a single momentum scale, the vertex form factors display in the IR the so-called ``zero crossing'', followed by a negative logarithmic divergence at the origin. However, particular care is needed with such a form factor away from the Landau gauge; in fact, when $\xi\neq0$, all $A_i$ terms in the BSE~\noeq{euclideanBS} are active, 
and, since such form factor would violate the condition of~\1eq{potdiv}, $B'_1(x)$ will be IR divergent. 

Thus, the behavior of $f_3$ will be modelled according to   
\begin{align}
f_3(z) = a\left[1 + b\ln\frac{z + m^2}{\mu^2}+c\ln\frac{z+ d\xi}{\mu^2}+ e\frac{m^2 (z-\mu^2)}{(z+m^2)(\mu^2+m^2)} \right],
\label{vertex3}
\end{align}
where $a=1.64$, $m=0.124\, \mbox{GeV}$, $b = e= -5.30$, $c = 5.40$,  and $d=0.005\,\mbox{GeV}^2$. Evidently, 
in the Landau gauge this expression displays the expected logarithmic IR divergence (and, in addition, reproduces well the lattice data of~\cite{Athenodorou:2016oyh}), whereas,  when $\xi\neq0$, only a zero crossing is present, and $f_3$ saturates at a finite value (left panel of~\fig{fig:bse_f3}). The corresponding solutions are 
shown in the right panel of the same figure; one observes that $B'_1$ is slightly suppressed with respect to the previous cases as a result of the large negative IR values acquired by $f_3$.   

\begin{figure}[!t]
\hspace{-0.5cm}\includegraphics[scale=0.61]{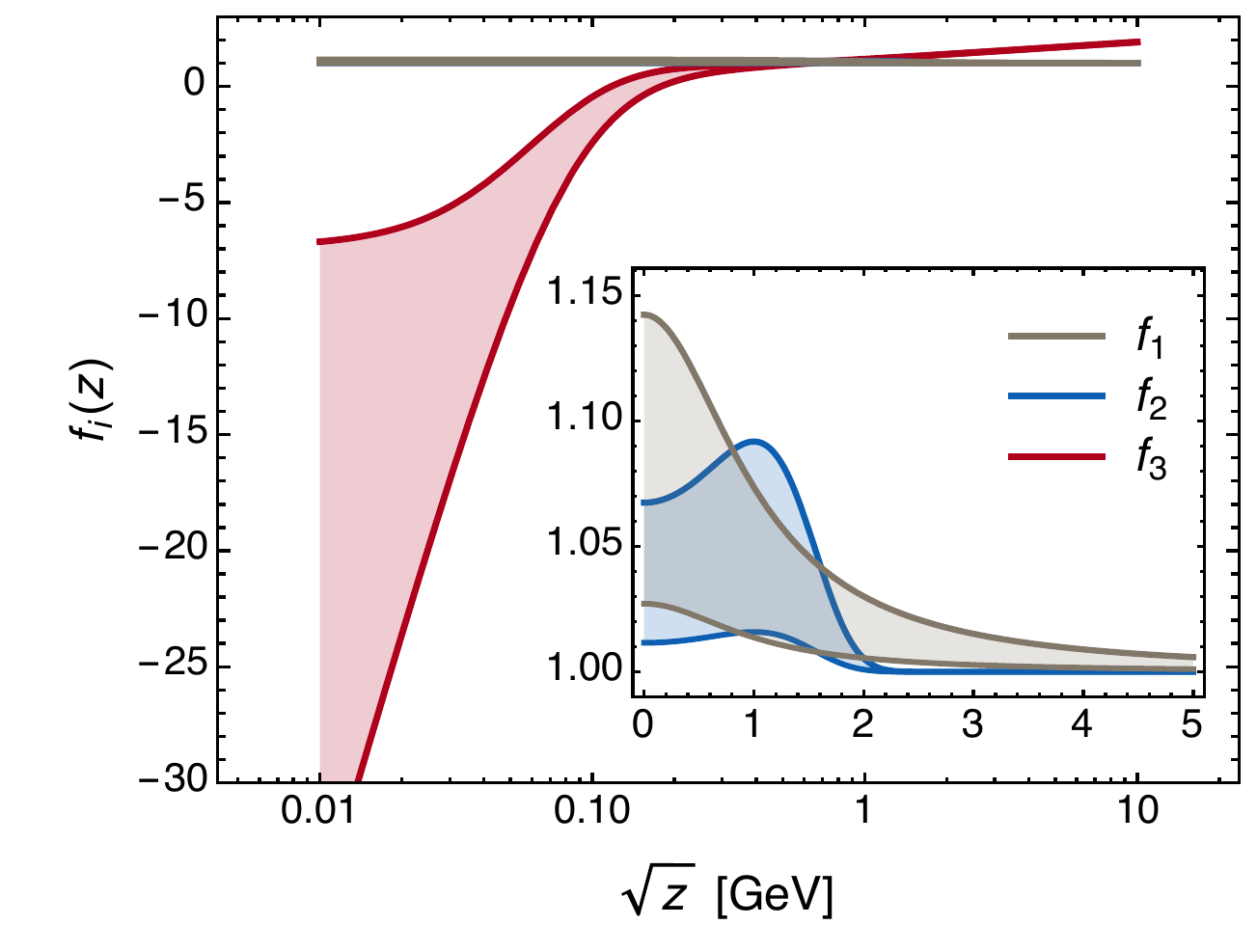}\hspace{.5cm}
\includegraphics[scale=0.61]{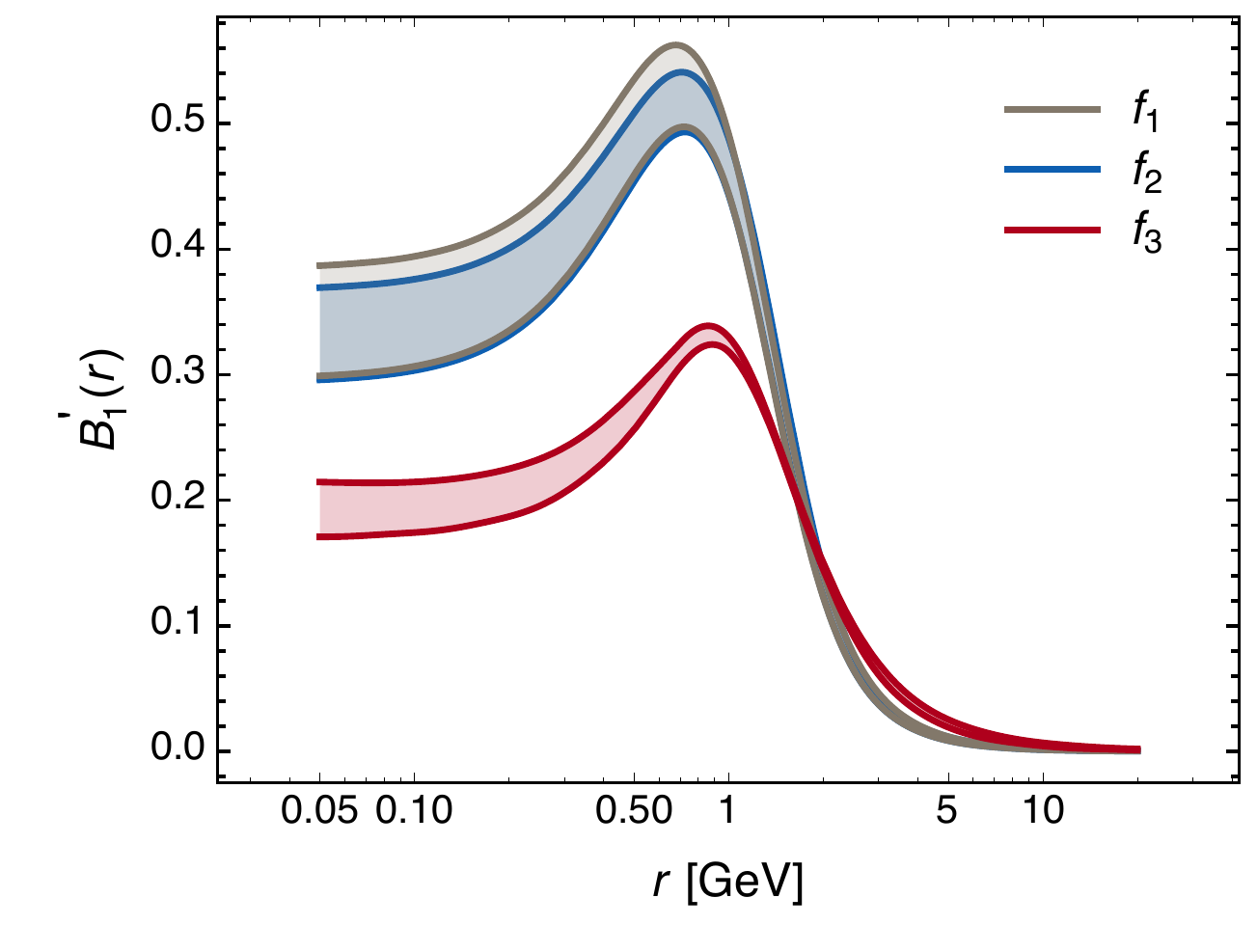}
\caption{\label{fig:comp} Variation of the three gluon vertex form factors $f_i(z)$ (left) and the corresponding BSE solutions $B_1^{\prime}(r)$ (right) when varying the gauge fixing parameter. Notice the relative small variations of the solutions when compared to the changes in the vertex form factors.}
\end{figure}

Finally, to better appreciate the reduced sensitivity of the BS solutions to variations in both the gauge fixing parameter as well as the form factors employed, in~\fig{fig:comp} we compare the $f_i$ with the corresponding $B_1^{\prime}$.

\section{\label{sec:glueball}Further Considerations}   

At first sight,  
the BSE depicted in Fig.~\ref{bse} appears to be identical to the
corresponding equation used in pure Yang-Mills theories to describe
the formation of a glueball out of two gluons (see, {\it e.g.},~\cite{Meyers:2012ka,Sanchis-Alepuz:2015hma}). This observation,
together with the fact that massless solutions have indeed been
found in our numerical treatment of the BSE, may raise the question of whether
a massless glueball could also exist in the physical spectrum
of the theory. However, 
no such state has ever been identified
in any of the studies presented in the literature~\cite{Morningstar:1999rf,Mathieu:2008me,Meyers:2012ka,Sanchis-Alepuz:2015hma}.

The way to resolve this apparent contradiction is by recognizing that,
despite their pictorial resemblance, the  BSE of the colored state and
that of the (color singlet) glueball are, in fact, rather different.

\begin{figure}[!t]
\includegraphics[scale=0.625]{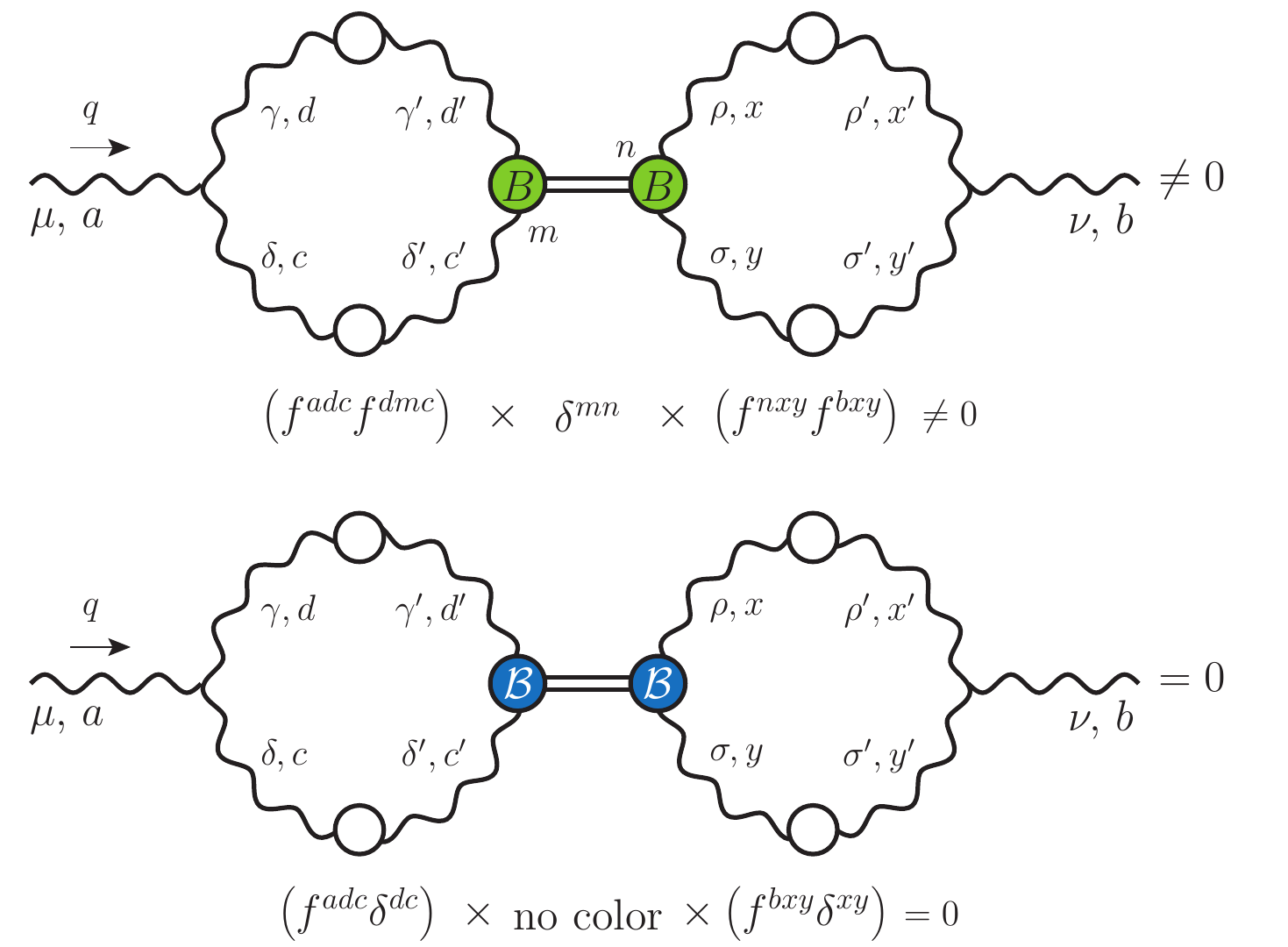}\caption{\label{fig:color} The {\it colored} bound states appearing in  the
nonperturbative three-gluon vertex (upper panel) and the vanishing 
contribution of the {\it color singlet bound state} (glueball) to the
gluon self-energy (lower panel).}
\end{figure}

To appreciate the reason for that, remember 
that the color structure of the amplitude describing 
the formation of  the massless colored state is given by  
\mbox{$B^{abc}_{\alpha\beta} (q,r,p) = f^{abc}B_{\alpha\beta} (q,r,p)$} [the $f^{abc}$ has been suppressed in the definition given in \1eq{UIB}]; evidently, the {\it colored} bound states form part of the nonperturbative three-gluon vertex, and appear inside the gluon self-energy though the typical diagram shown in the upper panel of~\fig{fig:color} (see also Fig.~5 in Ref.~\cite{Aguilar:2011xe}), leading to the infrared finiteness of the gluon propagator. On the other hand,  the corresponding color singlet amplitude describing the formation of a  glueball (no color) has the form \mbox{${\mathcal B}_{\alpha\beta}^{bc}(q,r,p) = \delta^{bc}{\mathcal B}_{\alpha\beta}(q,r,p)$}. Clearly, no such state could propagate inside the gluon propagator, as can be directly deduced from the diagram in the lower panel of~\fig{fig:color}. Note that the difference in the color structure affects the symmetry properties of $B_{\alpha\beta} (q,r,p)$ and ${\mathcal B}_{\alpha\beta}(q,r,p)$ under the exchange $\alpha \leftrightarrow \beta$ and $r \leftrightarrow p$, and, ultimately, their behavior at $q=0$. Specifically, the properties of \2eqs{BoseB}{B0} are a direct result of the full Bose symmetry of the vertex, and the fact that $f^{abc}=-f^{acb}$; instead, the Bose symmetry in the case of ${\mathcal B}_{\alpha\beta}^{bc}(q,r,p)$ yields the relation ${\mathcal B}_{\alpha\beta}(q,r,p) = {\mathcal B}_{\beta\alpha}(q,p,r)$, and, unlike $B_{\alpha\beta}(0,-p,p)$, the amplitude ${\mathcal B}_{\alpha\beta}(0,-p,p)$ does not have to vanish. 

It turns out that the aforementioned  differences in the color structures between the two amplitudes induce serious modifications in the form of the two BSE equations, which may therefore have entirely different types of solutions. In fact, whereas the BSE describing the pole formation is written in terms of $B_1^{\prime}$ [precisely due to \1eq{B0}], in the glueball case the BSE involves ${\mathcal B}_1$.  

To get an idea of the type of differences appearing in the glueball BSE, we may repeat some of the basic steps presented in Sect.~\ref{sec:bse}, now using the ${\mathcal B}_{\alpha\beta}^{mn}(q,r,p)$ instead of $B^{amn}_{\alpha\beta} (q,r,p)$. In particular, the equivalent of Eq.~\eqref{BSEq} reads (one may now set $q=0$ directly)
\begin{align}
{\mathcal B}_{\alpha\beta}^{mn}(0,r,-r)=  \int_k{\mathcal B}_{\gamma\delta}^{bc}(0,k,-k)\Delta^{\gamma\rho}(k)\Delta^{\delta\sigma}(k){\cal K}_{\rho\alpha\beta\sigma}^{bmnc}(-k,r,-r,k),
\label{BSEglue}
\end{align}
where the  lowest-order expansion of the four-gluon BS kernel ${\cal K}$ is also represented by the  diagrams $(b_1)$, $(b_2)$ and $(b_3)$  shown  in Fig.~\ref{4gkernel}. In doing that the the tree-level diagram $(b_1)$ reduces to 
\begin{align}
	(b_1)\to-iC_Ag^2\delta^{mn}\int_k\!{\mathcal B}_{\gamma\delta}(0,k,-k)\Delta^{\gamma\rho}(k)\Delta^{\delta\sigma}(k) (2g_{\rho\sigma}g_{\alpha\beta} -  g_{\rho\beta}g_{\alpha\sigma}-g_{\rho\alpha}g_{\beta\sigma}),
\end{align}
which is evidently symmetric under the exchange of indices $\alpha \leftrightarrow \beta$ contrary to what happens to Eq.~(\ref{b1diagram}). Thus, after the multiplication by the projector $P^{\alpha\beta}(r)$ the contribution from graph $(b_1)$ does {\it not} vanish. 

In addition, the diagrams $(b_2)$ and $(b_3)$ read
\begin{align}
	(b_2)&+(b_3)\to  \frac12g^2C_A\delta^{mn}\int_k\!{\mathcal B}_{\gamma\delta}(0,k,-k)\Delta^{\gamma\rho}(k)\Delta^{\delta\sigma}(k)  \\
	&\times\left[f^2(k-r)\Gamma^{(0)}_{\rho\mu\alpha}(-k,k-r,r)\Gamma^{(0)}_{\sigma\nu\beta}(k,r-k,-r)\Delta^{\mu\nu}(r-k)\right. \nonumber \\
	&\left.-f^2(k+r)\Gamma^{(0)}_{\rho\mu\beta}(-k,k+r,-r)\Gamma^{(0)}_{\sigma\nu\alpha}(k,-r-k,r)\Delta^{\mu\nu}(k+r)\right]\,, \nonumber
\end{align}
where the difference in the numerical factor with respect to Eq.~(\ref{b23diagram}) [$1/2$ instead of $1/4$] is due to the fact that distinct color identities need be employed in each case. Note finally that the term ${(r\!\cdot\!k)}/{r^2}$ would not appear in the equation equivalent to \1eq{bsemink}, given that the
origin of this term is the Taylor expansion of $B_1$ around $q=0$, which is not necessary now.

In summary, it should be clear from the above
discussion that, due to the differences in the
two dynamical equations induced by the color structures, 
one may not infer the formation of massless glueballs
from the corresponding formation of colored bound states.

\section{\label{sec:c}Conclusions}   

In the present work we have studied the dynamical formation of massless bound-state excitations by means of a homogeneous linear BSE, in order to establish the applicability of a particular gluon mass generating mechanism away from the Landau gauge. The BSE in question has been derived for general values of the gauge-fixing parameter, and was subjected to a number of simplifying assumptions that reduce its complexity. One of the main simplifications has been to retain only the form factor $B_1$, thus converting the would-be system of coupled BSEs into a single integral equation, whose kernel 
has been subsequently approximated by a dressed version of its lowest-order graphs.   
The main ingredients composing this kernel are 
the lattice data of~\cite{Bicudo:2015rma},  used for the gluon propagators,
and certain simple Ans\"atze that  model the ($\xi$-dependent) form factor, $f(z)$,  
of the three-gluon vertices.

The structure of the $\xi$-dependent terms contributing to the kernel reveals 
the need to distinguish between IR finite and IR divergent versions of $f(z)$.
When the $f(z)$ employed is IR finite, the 
detailed numerical study of this BSE reveals a continuity in the type of solutions 
that can be obtained as one varies $\xi$, which, in turn, may be interpreted as an 
indication that the departure from the Landau gauge is smooth and stable.
Instead, if one starts out with a logarithmically divergent $f(z)$, one may obtain a 
perfectly acceptable solution for the BSE in the Landau gauge, but this ceases to be true as 
$\xi$ departs from zero. The way around this problem was to postulate that, 
away from the Landau gauge, $f(z)$ turns IR finite, in which case one finds again 
well-behaved nontrivial solutions. 

This last point requires particular attention, because, if taken at face value, 
it would seem to suggest that if an independent calculation could 
conclusively establish that the IR divergent nature of $f(z)$  
persists away from the Landau gauge, then the mass generation mechanism  realized 
through the dynamical formation of massless excitations ought to be ruled out, 
or at least drastically revised. Conversely, one may say that 
if one accepts the lattice results of~\cite{Bicudo:2015rma} as valid, and 
attributes their origin to the aforementioned mechanism, then 
the IR finiteness of $f(z)$ seems to be an inescapable conclusion.
However, no such strong claims can be made at present, given the 
approximate nature of the kernel of the BSE, coupled with the fact that 
certain terms [\eg terms of type \n{ii$_{\,b}$}] have been discarded for the sake of algebraic expedience.  
   
Given the above discussion,  
it would certainly be interesting to inquire what kind 
of mechanism might make $f(z)$ IR finite away from the Landau gauge. 
To that end, let us first recall~\cite{Aguilar:2013vaa} that the Slavnov-Taylor identity of the three-gluon vertex 
relates, in a rather intricate way, the behaviour of $f(z)$ to that of the
ghost propagator, $D(q^2)$, whose dressing function, $F(q^2) = q^2D(q^2)$, is IR finite 
in the Landau gauge~\cite{Bogolubsky:2009dc,Bogolubsky:2007ud}.
If one were to move away from the Landau gauge, the aforementioned connection between these two quantities is likely to persist, given that the Slavnov-Taylor identity maintains its form for all values of $\xi$; however, as has been shown in~\cite{Aguilar:2015nqa}, away from the Landau gauge a qualitative change takes place in the IR behavior of the ghost dressing functions, which, instead of saturating to a constant value,  approaches to zero very slowly~(similar conclusions were also reached in~\cite{Huber:2015ria}). It is therefore plausible that this particular difference in the behavior of the ghost dressing function might eventually account for the corresponding change in $f(z)$.

There is a number of potential improvements that could strengthen the conclusions 
of this preliminary exploration, at the expense of adding various layers of technical 
complexity to the problem at hand. For instance, one may retain the form factor ${\overline B}_3$ in~\1eq{I1f}, 
and attempt to construct a system of coupled BSE involving both $B_1$ and ${\overline B}_3$.
In addition, one may introduce vertex form factors 
with more complicated momentum dependence, given that the $f$ is in reality a function 
of three kinematic variables, rather than a single one, as was supposed for simplicity here.

A rather interesting possibility would be to refrain from using the lattice data for the gluon propagators as input to the BSE, and study instead the system of equations formed when the BSE is {\it coupled} to the SDE of the gluon propagator, given in~\fig{QQ-SDE}. One advantage of such an approach is that it would allow, at least in principle, to extend the analysis to values of $\xi>0.5$, and in particular reach the Feynman gauge, $\xi=1$.  From the technical point of view, this endeavor would entail the explicit derivation of the gluon SDE for a general $\xi$, a task that is still pending, at least within the framework employed in~\cite{Aguilar:2011xe,Ibanez:2012zk,Aguilar:2016vin}.

\acknowledgments 

The authors thank J.~Rodriguez-Quintero for useful discussions. The research of J.~P. is supported by the Spanish MEYC under  FPA2014-53631-C2-1-P and SEV-2014-0398, and Generalitat Valenciana under grant Prometeo~II/2014/066. The work of  A.~C.~A.  is supported by the National Council for Scientific and Technological Development - CNPq under the grant 305815/2015.


\end{document}